\documentclass[letterpaper,10pt,twocolumn,twoside, conference]{IEEEtran}
\IEEEoverridecommandlockouts                    
\usepackage{times}
\usepackage{setspace}
\usepackage{url}
\spacing{1}
\usepackage[utf8]{inputenc}
\usepackage[T1]{fontenc}
\usepackage{graphicx}		
\usepackage{wrapfig}
\usepackage[format=plain,font=footnotesize,labelfont=bf,labelsep=period]{caption}
\usepackage[export]{adjustbox}
\usepackage{float}
\usepackage{dblfloatfix}

\usepackage{amsmath} 
\usepackage{amssymb}  
\usepackage{amsthm}
\usepackage{mathtools}
\usepackage[normalem]{ulem}
\usepackage{paralist}	
\usepackage[space]{grffile} 
\usepackage{xcolor}
\usepackage{bbm}
\usepackage{placeins} 
\usepackage{array}
\usepackage{siunitx}
\usepackage{hyperref}
\usepackage{cleveref}
\usepackage{bm} 
\usepackage{algorithm}
\usepackage{algpseudocode}
\usepackage{cleveref}
\usepackage{cite}

\newcommand{\red}[1]{{\color{red} #1}}

\newtheorem{theorem}{Theorem}
\newtheorem{corollary}{Corollary}
\newtheorem{lemma}{Lemma}
\theoremstyle{definition}
\newtheorem{definition}{Definition}
\theoremstyle{remark}
\newtheorem{remark}{Remark}
\theoremstyle{definition}
\newtheorem{assumption}{Assumption}
\theoremstyle{definition}

\theoremstyle{definition}

\theoremstyle{definition}
\newtheorem{fact}{Fact}
\theoremstyle{definition}

\theoremstyle{definition}

\renewcommand{\cal}[1]{\mathcal{ #1 }}
\newcommand{\mb}[1]{\mathbf{ #1 }}
\newcommand{\bs}[1]{\boldsymbol{#1}}

\newcommand{\R}{\mathbb{R}}

\newcommand{\E}{\cal{E}} 
\newcommand{\BEZc}{r} 
\newcommand{\BEZC}{\mb{r}}
\newcommand{\bez}{\xi}
\newcommand{\Bez}{\bs \xi}
\newcommand{\BEZ}{\bs{\zeta}}
\newcommand{\z}{\mb{z}} 
\newcommand{\RobustInvariant}{\Omega}
\newcommand{\ErrorSet}{\mathcal{E}}

\newcommand{\x}{\mb{x}} 
\newcommand{\state}{\x} 

\newcommand{\f}{\mb{f}}
\newcommand{\g}{\mb{g}}
\newcommand{\A}{\mb{A}}
\newcommand{\B}{\mb{B}}
\newcommand{\C}{\mb{C}}

\newcommand{\grad}{\nabla}

\DeclareMathOperator*{\argmin}{argmin}

\newcommand{\tintc}{[\underline{t},\overline{t}]}
\newcommand{\tinto}{[\underline{t},\overline{t})}

\newcommand{\mpcindex}{i}

\title{\LARGE \textbf{Multi-Rate Planning and Control of Uncertain Nonlinear Systems: \\ Model Predictive Control and Control Lyapunov Functions}}
\author{Noel Csomay-Shanklin$^*$, Andrew J. Taylor$^*$, Ugo Rosolia, Aaron D. Ames 
\thanks{This work is supported by the National Science foundation (CPS Award \#1932091, NRI Award \#1924526,  CMMI Award \#1923239)), and the AFOSR Test and Evaluation Program (FA9550-19-1-0302).}
\thanks{$^*$ Authors contributed equally. N. Csomay-Shanklin, A. J. Taylor, U. Rosolia, and A. D. Ames are with the Department of Computing and Mathematical Sciences, California Institute of Technology, Pasadena, CA 91125, USA, {\tt\small \{noelcs, ajtaylor, urosolia, ames\}@caltech.edu}.}
}

\begin{document}

\maketitle
\begin{abstract}
Modern control systems must operate in increasingly complex environments subject to safety constraints and input limits, and are often implemented in a hierarchical fashion with different controllers running at multiple time scales. 
Yet traditional constructive methods for nonlinear controller synthesis typically ``flatten'' this hierarchy, focusing on a single time scale, and thereby limited the ability to make rigorous guarantees on constraint satisfaction that hold for the entire system.  
In this work we seek to address the stabilization of constrained nonlinear systems through a \textit{multi-rate} control architecture. This is accomplished by iteratively planning continuous reference trajectories for a nonlinear system using a linearized model and Model Predictive Control (MPC), and tracking said trajectories using the full-order nonlinear model and Control Lyapunov Functions (CLFs). Connecting these two levels of control design in a way that ensures constraint satisfaction is achieved through the use of \textit{B\'{e}zier curves}, which enable planning continuous trajectories respecting constraints by planning a sequence of discrete points. Our framework is encoded via convex optimization problems which may be efficiently solved, as demonstrated in simulation.

\end{abstract}

\newif\ifextended 
\extendedfalse 
%

\section{Introduction}



The study and design of nonlinear control systems has long been framed through the lens of stabilization, often in an optimal sense. This is coupled with the fact that one typically considers a single model, implicitly representing a single time scale. However in most modern engineering settings, especially in the context of autonomous and robotic systems, the task of stabilization is complicated by the need to meet safety-critical constraints on the system's state while respecting input limitations. To address this need, implementations often utilize a hierarchical approach that spans multiple time-scales, from the planning layer---which typically leverages discrete-time models---to the real-time controller layer which often considers continuous-time representations. Thus it is necessary to develop efficient control synthesis techniques that provide rigorous guarantees of stability, even in the presence of such constraints, and across multiple time scales. 

\par
At the level of real-time control design, a rich catalog of methods have been developed for stabilizing nonlinear systems in the presence of unknown disturbances by utilizing underlying structural properties of the system \cite{isidori1995nonlinear, kokotovic2001constructive, khalil2002nonlinear, sepulchre2012constructive}. In particular, the tools of Control Lyapunov Functions (CLFs) \cite{artstein1983stabilization, sontag1989universal} and Input-to-State Stability \cite{Sontag1989smooth} have enabled the joint synthesis of stabilizing controllers and Lyapunov certificates of stability in the presence of disturbances, including through convex optimization \cite{freeman1996robust, ames2013towards, kolathaya2018iss}. These methods for stabilization yield highly structured controllers, and modifying these designs to accommodate state and input constraints may destroy the stability properties guaranteed by the controller. This issue is often circumnavigated theoretically by limiting the domain on which stability is guaranteed, effectively ignoring constraints. 

\par
In contrast, Model Predictive Control (MPC) provides an effective method for addressing constraints \cite{allgower2004nonlinear,allgower2012nonlinear, borrelli2017predictive}. This is achieved by directly incorporating constraints a into controller that iteratively plans a finite sequence of states and inputs that are related through a discrete model of the system dynamics and satisfy required constraints. Although MPC has been successfully demonstrated in several challenging control settings \cite{dicarlo2018dynamic, sleiman2021unified,falcone2007predictive, hrovat2012development, lima2018experimental, bengea2012model, serale2018model, maddalena2020data, rosolia2019learning, liniger2015optimization, hewing2019cautious}, it is rarely implemented in real-time using the full-order continuous time nonlinear dynamics while accounting for unknown disturbances acting on the system. Thus, MPC implementation for nonlinear systems usually lack strong theoretical guarantees on constraint satisfaction in the presence of disturbances. This is because (i) it is typically difficult to find a closed-form expression for the exact temporal discretization of continuous time nonlinear dynamics \cite{nevsic1999sufficient}, (ii) approximating the exact discretization through numerical integration typically yields a non-convex relationship between planned states and inputs, and (iii) exactly propagating disturbances through high-dimensional nonlinear dynamics is often computationally intractable~\cite{bansal2017hamilton}. These challenges often preclude achieving the computational efficiency needed for real-time implementation.

\begin{figure}
    \centering
    \includegraphics[width=0.85\linewidth]{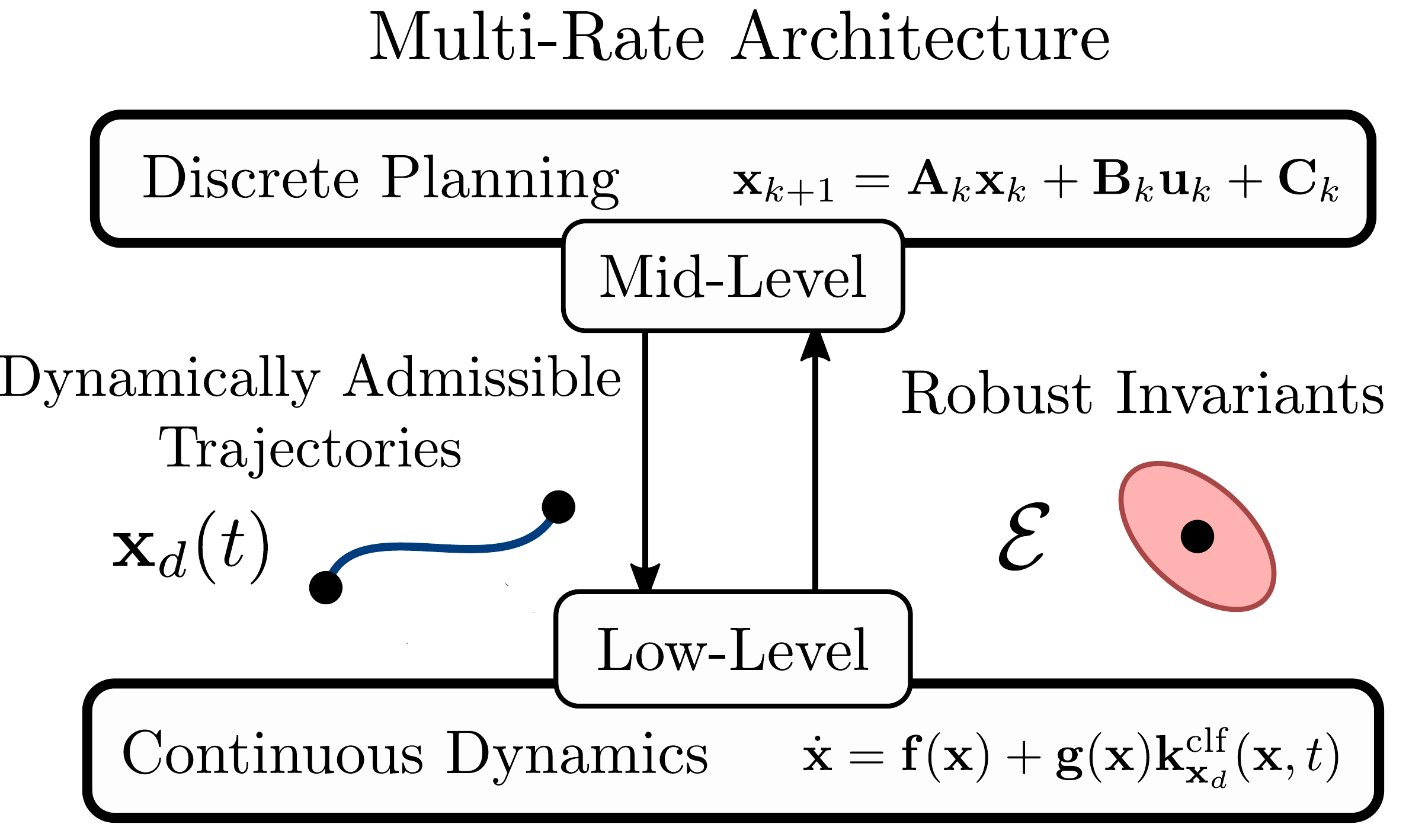}
    \caption{Overview of Multi-Rate Architecture, with discrete planning producing reference trajectories at a mid-level and continuous controllers producing invariant sets at a low-level.}
    \vspace{-5mm}
    \label{fig:my_label}
\end{figure}

\par

The difficulty in realizing MPC based controllers at a fast enough rate to allow for real-time implementation 
is often resolved by using an approximate model of the system dynamics that is amenable to efficient planning, typically through reduced-order models or via linearization and temporal discretization of the continuous time nonlinear system dynamics \cite{borrelli2017predictive, rosolia2019learning, liniger2015optimization,carvalho2013predictive, hewing2019cautious}. The use of such approximations creates a gap between the system which is being planned for and the actual evolution of the nonlinear system, requiring an additional measure of robustness to ensure constraint satisfaction. This robustness is often achieved by tightening the constraint sets by the maximum deviation between the approximate model and the continuous time nonlinear system dynamics \cite{gao2014tube, kogel2015discrete, yu2013tube, singh2017robust,kohler2020computationally,herbert2017fastrack, singh2018robust,yin2020optimization}. Approximating worse-case deviations is typically done using properties of the dynamics which may be difficult to compute, such as Lipschitz constants for which over-approximations yield conservativeness, or by solving computationally intensive optimization programs. More recently, hierarchical control frameworks have been proposed that plan with an approximate model, but directly address nonlinear dynamics with a low-level controller \cite{rosolia2020unified, rosolia2021multirate}. However, this work does not address if the low-level controller respects state and input constraints as it follows the planned trajectory under disturbances.


%
%
\par
In this work we propose a novel multi-rate control architecture that unifies the planning capabilities of Model Predictive Control with the ability to directly address nonlinear dynamics provided by Control Lyapunov Functions. The fundamental tool that allows our framework to explicitly address the relationship between a planner and controller operating at different time scales are \textit{B\'{e}zier curves} \cite{kamermans2020primer, farin2002curves}. By directly planning over the \textit{control points} that parameterize B\'{e}zier curves, we capitalize on a critical convex hull property to ensure that state and input constraints are met by the nonlinear system evolving under an optimization-based CLF controller. While B\'{e}zier curves have been used in motion planning, or to verify constraint satisfaction after solving an MPC problem \cite{muehlebach2017implementation}, this is to the best of our knowledge the first result directly planning over B\'{e}zier control points in an MPC formulation, and using the resulting continuous trajectories to ensure constraint satisfaction for a nonlinear system with disturbances.


We begin in Section \ref{sec:lowlevel} by reviewing nonlinear dynamics, and how structural properties can be used to synthesize CLFs and optimization-based controllers for stabilizing a class of \textit{dynamically admissible} reference trajectories. These controllers yield a description of how accurately a reference trajectory is tracked in the presence of disturbances that is amenable to being incorporated into planning. Next, in Section \ref{sec:bezier} we provide a review of B\'{e}zier curves, and show how they may be used to synthesize reference trajectories for the disturbed nonlinear system such that state constraints are satisfied.  Section \ref{sec:input} uses the properties of B\'{e}zier curves in conjunction with the structure of the low-level controller to formulate constraints on B\'{e}zier control points that ensure the low-level controller satisfies input constraints. In Section \ref{sec:multirate} we integrate the preceding constructions into an MPC formulation that plans over B\'{e}zier control points and synthesizes continuous reference trajectories using a locally linearized and discretized model while ensuring recursive feasibility. We conclude in Section \ref{sec:sim} with simulation results.
We note that proofs may be found in the appendix.

\section{Low-Level Controller Design}
\label{sec:lowlevel}
In this section we review nonlinear dynamical systems and discuss the design of nonlinear feedback controllers that provide a measure of disturbance rejection. Importantly, these controllers will yield a quantitative description of reference trajectory tracking that is amenable to being directly incorporated into the synthesis of the reference trajectory itself.

Consider the nonlinear control-affine system:
\begin{equation}
    \label{eqn:openloop_state}
    \dot{\mb{x}} = \underbrace{\begin{bmatrix} \mb 0&\mb I\\0&\mb{0}^\top\end{bmatrix}\x+\begin{bmatrix}\mb 0\\ f(\mb{x}) \end{bmatrix}}_{ \mb{f}(\mb{x})}+\underbrace{\begin{bmatrix} \mb 0\\ g(\mb{x})\end{bmatrix}}_{\mb{g}(\mb{x})}u + \mb{w}(t),
\end{equation}
with state $\mb{x}\in\R^n$, input $u\in\R$, piecewise continuous\footnote{This definition is taken as in \cite{khalil2002nonlinear}, with piecewise continuity requiring the existence of one-sided limits at points of discontinuity.} disturbance signal $\mb{w}:\R_{\geq 0}\to\R^n$, and functions $f:\R^n\to\R$ and $g:\R^n\to\R$, assumed to be continuously differentiable on $\R^n$. Furthermore, we make the following assumption:
\begin{assumption}
The function $f$ satisfies $f(\mb{0}) = 0$ and the function $g$ satisfies $g(\state)\ne0$ for all $\x\in\R^n$.
\end{assumption}
\noindent The first assumption takes the origin to be an unforced equilibrium point of the undisturbed system. The second assumption amounts to the system \eqref{eqn:openloop_state} possessing a relative degree \cite{isidori1995nonlinear}. We note that while we consider a single-input, single-output system, this is purely to simplify the presentation of our contributions, and the subsequent developments may be easily extended to the multiple-input, multiple-output setting under an equivalent assumption of a vector relative degree.

Let $\underline{t},\overline{t}\in\R_{\geq 0}$ with $\underline{t}<\overline{t}$, and let $k:\R^n\times\tintc\to\R$ be a feedback controller that is locally Lipschitz continuous with respect to its first argument\footnote{This definition is taken as in \cite{khalil2002nonlinear}, with local Lipschitz continuity holding with a Lipschitz constant that is uniform in the function's second argument.} and piecewise continuous with respect to its second argument on $\R^n\times\tintc$. This controller yields the closed-loop system:
\begin{equation}
    \label{eqn:closedloop_state}
    \dot{\mb{x}} = \mb{f}(\mb{x})+\mb{g}(\mb{x})k(\mb{x},t)+\mb{w}(t).
\end{equation}
As the functions $\mb{f}$, $\mb{g}$, and $k$ are locally Lipschitz continuous with respect to $\mb{x}$ and $k$ is piecewise continuous with respect to $t$, for any initial condition $\mb{x}_0 \in \R^n$ and any piecewise continuous disturbance $\mb{w}:\R_{\geq 0}\to\R^n$, there exists an interval $I(\underline{t},\mb{x}_0,\mb{w})\triangleq[\underline{t},\underline{t}+\delta(\mb{x}_0,\mb{w}))$ with $ \delta(\mb{x}_0,\mb{w})\in\R_{>0}$ such that the system \eqref{eqn:closedloop_state} has a unique piecewise continuously differentiable\footnote{Piecewise continuous differentiability is taken to mean a continuous function with a derivative defined on the open intervals of a finite partition with one-sided limits.} solution $\bs{\varphi}:I(\underline{t},\mb{x}_0,\mb{w})\to\R^n$ satisfying:
\begin{align}
    \dot{\bs{\varphi}}(t) & = \mb{f}(\bs{\varphi}(t)) + \mb{g}(\bs{\varphi}(t))k(\bs{\varphi}(t),t)+\mb{w}(t), \label{eqn:soldiff_state}\\
    \bs{\varphi}(\underline{t}) & = \mb{x}_0, \label{eqn:solic_state}
\end{align}
for almost all $t\in I(\underline{t},\mb{x}_0,\mb{w})$ \cite{khalil2002nonlinear}. 

With a view towards controller design, the system \eqref{eqn:openloop_state} may also be used to define a class of reference trajectories:
\begin{definition}[\textit{Dynamically Admissible Trajectory}]
A piecewise continuously differentiable function $\x_d:[\underline{t},\overline{t}]\to\R^n$ is a \textit{dynamically admissible trajectory} for the system \eqref{eqn:openloop_state} if there is a piecewise continuous function $u_d:[\underline{t},\overline{t}]\to\R$ such that:
\begin{equation}
    \label{eqn:dynadmistraj}
    \dot{\x}_d(t) = \f(\x_d(t))+\g(\x_d(t))u_d(t),
\end{equation}
for almost all $t\in[\underline{t},\overline{t}]$.
\end{definition}
Given a dynamically admissible trajectory $\mb{x}_d:[\underline{t},\overline{t}]\to\R^n$ for \eqref{eqn:openloop_state}, let us denote:
$
    \dot{\mb{x}}_d(t) = \begin{bmatrix} \dot{x}_{d}^1(t) & \cdots & \dot{x}_{d}^n(t) \end{bmatrix}^\top,
$
and define a error function $\mb{e}_{\mb{x}_d}:\R^n\times\tintc\to\R^n$: 
\begin{equation}
    \label{eqn:errorfunc}
    \mb{e}_{\mb{x}_d}(\mb{x},t) = \mb{x}-\mb{x}_d(t),
\end{equation}
and its derivative $\dot{\mb{e}}_{\mb{x}_d}:\R^n\times\tintc\times\R\to\R^n$ as:
\begin{align}
    \label{eqn:errorderivfunc}
    \dot{\mb{e}}_{\mb{x}_d}(\mb{x},t,u) = &~\mb{f}(\mb{x})+\mb{g}(\mb{x})u + \mb{w}(t)-\dot{\mb{x}}_d(t).
\end{align}
Denoting:
\begin{equation}
\label{eqn:calFxd}
    \mathcal{F}_{\mb{x}_d}(\mb{x},t) = f(\mb{x})-\dot{x}_d^n(t),
\end{equation}
the structure of the system \eqref{eqn:openloop_state} implies that:
\begin{align}
\label{eqn:errordotdecomp}
    \dot{\mb{e}}_{\mb{x}_d}(\mb{\x},t,u) = & \overbrace{\begin{bmatrix}\mb{0} & \mb{I} \\ 0 & \mb{0}^\top  \end{bmatrix}\mb{e}_{\mb{x}_d}(\mb{x},t)  + \begin{bmatrix} \mb{0} \\ \mathcal{F}_{\mb{x}_d}(\mb{x},t)\end{bmatrix}}^{\mb{f}_{\mb{x}_d}(\mb{x},t)} \\ & \hspace{27.8mm}+ \mb{g}(\mb{x})u + \mb{w}(t). \nonumber
\end{align}
This structure in conjunction with the assumption that $g(\mb{x})\neq 0$ for any $\mb{x}\in\R^n$
enables a controller $k^{\textrm{fbl}}_{{\mb{x}_d}}:\R^n\times\tintc\to\R$:
\begin{equation}
\label{eqn:FBL_CL_gain}
    k^{\rm fbl}_{\mb{x}_d}(\mb{x},t) = g(\mb{x})^{-1}\left(-\mathcal{F}_{\mb{x}_d}(\mb{x},t)-\mb{K}^\top\mb{e}_{\mb{x}_d}(\mb{x},t)\right),
\end{equation}
where $\mb{K}\in\R^{n}$ is selected to yield the relationship:
\begin{equation}
 \dot{\mb{e}}_{\mb{x}_d}(\mb{x},t,k_{\rm fbl}(\mb{x},t)) =  {\mb{F}}\mb{e}_{\mb{x}_d}(\mb{x},t) + \mb{w}(t), \label{eqn:errordotcloop}
\end{equation}
with $\mb{F}\in\R^{n\times n}$ a Hurwitz matrix. For any $\mb{Q}\in\mathbb{S}^n_{\succ 0}$ (symmetric positive definite matrices) there exists a unique $\mb{P}\in\mathbb{S}^n_{\succ 0}$ solving the Continuous Time Lyapunov Equation:
\begin{equation}
    \mb{F}^\top\mb{P}+\mb{PF} = -\mb{Q}.
\end{equation}
For a particular $\mb{Q}$, the corresponding solution $\mb{P}$ may be used to define the following function $V_{\mb{x}_d}:\R^n\times\tintc\to\R_{\geq 0}$:
\begin{equation}
\label{eqn:lyap}
    V_{\mb{x}_d}(\mb{x},t) = \mb{e}_{\mb{x}_d}(\mb{x},t)^\top\mb{P}\mb{e}_{\mb{x}_d}(\mb{x},t).
\end{equation}
Denoting $\grad V_{\mb{x}_d}(\mb{x},t) = 2\mb{e}_{\mb{x}_d}(\mb{x},t)^\top\mb{P}$, we have that:
\begingroup\makeatletter\def\f@size{9}\check@mathfonts
\begin{align}
   &\lambda_{\textrm{min}}(\mb{P})\Vert\mb{e}_{\mb{x}_d}(\mb{x},t)\Vert_2^2 \leq  V_{\mb{x}_d}(\mb{x},t)   \leq  ~\lambda_{\textrm{max}}(\mb{P})\Vert\mb{e}_{\mb{x}_d}(\mb{x},t)\Vert_2^2, \label{eqn:lyapbounds} \\ 
&\grad V_{\mb{x}_d}(\mb{x},t)(\mb{f}_{\mb{x}_d}(\mb{x},t)+\mb{g}(\mb{x})k_{\mb{x}_d}^{\rm fbl}(\mb{x},t)) \label{eqn:lyapdecay} \\ & \qquad\qquad\qquad\qquad\qquad\qquad \leq   -\lambda_{\rm min}(\mb{Q})\Vert\mb{e}_{\mb{x}_d}(\mb{x},t)\Vert_2^2, \nonumber
\end{align}
\endgroup
for all $\mb{x}\in\R^n$ and $t\in\tintc$. Let $\gamma = 4\lambda_{\max}(\mb{P})^3/\lambda_{\rm min}(\mb{Q})^2$ and for a given disturbance signal $\mb{w}:\R_{\geq 0}\to\R^n$ define $\Vert\mb{w}\Vert_\infty = \sup_{t\geq 0}\Vert\mb{w}(t)\Vert_2$. The preceding construction yields the following result:
\begin{lemma}
\label{lem:lowleveliss}
Let $\overline{w}\in\R_{\geq 0}$, and for $t\in\tintc$ define the set:
    \begin{equation}
    \label{eqn:robustInvariant}
        \RobustInvariant_{\mb{x}_d}(t, \overline{w}) = \{\mb{x}\in\R^n ~|~ V_{\mb{x}_d}(\mb{x},t) \leq \gamma\overline{w}^2 \}.
    \end{equation}
Let the controller $k:\R^n\times\tintc\to\R$ satisfy: 
\begin{align}
\label{eqn:cntrllyapreq}
    & \grad V_{\mb{x}_d}(\mb{x},t)(\mb{f}_{\mb{x}_d}(\mb{x},t)+\mb{g}(\mb{x})k(\mb{x},t)) \\ & \qquad\qquad\qquad\qquad\qquad\qquad \leq -\lambda_{\min}(\mb{Q})\Vert\mb{e}_{\mb{x}_d}(\mb{x},t)\Vert^2_2, \nonumber
\end{align}
for almost all $t\in\tintc$ and all $\mb{x}\in\RobustInvariant_{\mb{x}_d}(t,\overline{w})$. Then for initial time $\underline{t}$, any initial condition $\mb{x}_0\in\RobustInvariant_{\mb{x}_d}(\underline{t},\overline{w})$, and any disturbance signal $\mb{w}$ satisfying $\Vert\mb{w}\Vert_{\infty}\leq \overline{w}$, we have that $I(\underline{t},\mb{x}_0,\mb{w}) = \tinto$, and $\bs{\varphi}(t) \in \RobustInvariant_{\mb{x}_d}(t,\overline{w})$ for all $t\in[\underline{t},\overline{t})$, and $\lim_{t\to\overline{t}}\bs{\varphi}(t)$ exists and satisfies $\lim_{t\to\overline{t}}\bs{\varphi}(t)\in\RobustInvariant_{\mb{x}_d}(\overline{t},\overline{w})$.
\end{lemma}
The preceding result follows by a standard input-to-state stability argument \cite{Sontag1989smooth}. For any $t\in\tintc$, the set $\RobustInvariant_{\mb{x}_d}(t,\overline{w})$ captures how accurately the nonlinear closed-loop system \eqref{eqn:closedloop_state} tracks $\mb{x}_d$ with disturbances. Importantly, for a given $t\in\tintc$ the set $\RobustInvariant_{\mb{x}_d}(t,\overline{w})$ is convex -- as we will see later, this property will allow us to efficiently synthesize a dynamically admissible trajectory $\mb{x}_d$ while knowing how accurately it will be tracked and ensuring state and input constraint satisfaction.

In contrast to cancelling the nonlinear dynamics to achieve linear dynamics as in \eqref{eqn:errordotcloop}, which may be unnecessary and inefficient \cite{freeman1996robust}, Control Lyapunov Functions (CLFs) provide an alternative method for synthesizing stabilizing controllers via convex optimization. In particular, we have that \eqref{eqn:lyapdecay} implies:
\begin{align}
\label{eqn:clfdecay}
&\inf_{u\in\R}\grad V_{\mb{x}_d}(\mb{x},t)(\mb{f}_{\mb{x}_d}(\mb{x},t)+\mb{g}(\mb{x})u) \\ & \qquad\qquad\qquad\qquad\qquad\qquad \leq  -\lambda_{\min}(\mb{Q})\Vert\mb{e}_{\mb{x}_d}(\mb{x},t)\Vert^2_2. \nonumber
\end{align}
for all $\mb{x}\in\R^n$ and $t\in\tintc$. Define a feed-forward controller $k_{\mb{x}_d}^{\rm ff}:\R^n\times\tintc\to\R$ as:
\begin{equation}
\label{eqn:feedforward}
    k_{\mb{x}_d}^{\textrm{ff}}(\mb{x},t) = -g(\mb{x})^{-1}\mathcal{F}_{\mb{x}_d}(\mb{x},t).
\end{equation}
This feed-forward controller is incorporated into the following controller specified via a convex quadratic program (QP):
\begin{align}
\tag{CLF-QP} \label{clf-qp}
    & k_{\mb{x}_d}^\textrm{clf}(\mb{x},t) =  \argmin_{u\in\R} \frac{1}{2}\Vert u-k_{\mb{x}_d}^{\textrm{ff}}(\mb{x},t)\Vert_2^2 \\ & \textrm{s.t.}  ~ \grad V_{\mb{x}_d}(\mb{x},t)(\mb{f}_{\mb{x}_d}(\mb{x},t)+\mb{g}(\mb{x})u) \leq  -\lambda_{\min}(\mb{Q})\Vert\mb{e}_{\mb{x}_d}(\mb{x},t)\Vert^2_2. \nonumber
\end{align}
Note that the constraint in this controller ensures that $k_{\mb{x}_d}^{\rm clf}$ satisfies the condition in \eqref{eqn:cntrllyapreq}.

\section{B\'{e}zier Curves \& State Constraints}
\label{sec:bezier}


In this section we present the first main contribution of this work by addressing how the properties of the low-level tracking controller can be used to place requirements on a dynamically admissible trajectory $\mb{x}_d$ that ensure state constraint satisfaction by the closed-loop nonlinear system \eqref{eqn:closedloop_state} evolving under controllers such as $k_{\mb{x}_d}^{\rm fbl}$ or $k_{\mb{x}_d}^{\rm clf}$. 

We first make the following assumption regarding the state constraints for the system:
\begin{assumption}
\label{ass:statepolytope}
The state constraint set $\mathcal{X}\subset\mathbb{R}^n$ is a compact, convex polytope, with the existence of $\mb{L}_j\in\R^n$ and $\ell_j\in\R$ for $j=1,\ldots, q$ such that $\mathcal{X} = \{\x \in \R^n ~|~ \forall j,\, \mb L^\top_j\x \leq \ell_j\}$. Furthermore, we have that $\mb{0}\in\textrm{Int}(\mathcal{X})$.
\end{assumption}
\noindent Given the above state constraints, it is not the case -- even for a dynamically admissible trajectory satisfying $\mb{x}_d(t)\in\mathcal{X}$ for all $t\in\tintc$ -- that the state will remain inside the set $\mathcal{X}$, as we may have that $\RobustInvariant_{\mb{x}_d}(t,\overline{w})\nsubseteq\mathcal{X}$ for some $t\in\tintc$. To ensure these constraints are met by the closed-loop system without directly modifying the low-level control design, we will incorporate information about the low-level controller when constructing $\x_d$. The core tool that will enable incorporating this information is \emph{B\'{e}zier curves} \cite{kamermans2020primer}.

Let $T\in\R_{>0}$. A B\'ezier curve  $\BEZc:[0,T]\to\mathbb{R}$ of order $p$ is defined as:
\begin{align}
\label{eqn:bez}
    \BEZc(\tau) = \Bez_0^\top \z(\tau),
\end{align}
where $\Bez_0 = \begin{bmatrix}\bez_{0,0} & \ldots & \bez_{0,p}\end{bmatrix}^\top\in\mathbb{R}^{p+1}$ is a vector with elements consisting of the $p+1$ \textit{control points}, $\bez_{0,i}\in\mathbb{R}$, of the curve and $\z:[0,T]\to\mathbb{R}^{p+1}$ is a Bernstein polynomial defined elementwise as:
\begin{align}
    z_{i}(\tau) = \binom{p}{i} \left(\frac{\tau}{T}\right)^i\left(1-\frac{\tau}{T}\right)^{p-i}, ~~ i=0,\dots, p.
\end{align}
The curve $\BEZc$ is smooth, and there exists a matrix\footnote{The matrices $\mb{H}$ and $\mb{D}$ are uniquely defined by the order of the B\'ezier curve $p$ and can be constructed as shown in appendix.\label{HandDfootnote}} $\mb{H}\in\R^{p+1\times p+1}$ such that the $j^{th}$ derivative of $\BEZc$ is given by:
\begin{equation}
\label{eqn:bezjderiv}
    {\BEZc}^{(j)}(\tau) = \frac{1}{T^j}\Bez_0^\top \mb H^j \z (\tau) \triangleq \bs{\xi}_j^\top\mb{z}(\tau).
\end{equation}
Consequently, $\BEZc^{(j)}:[0,T]\to\R$ is a B\'ezier curve of order $p$ with the elements of $\Bez_j$ (which are uniquely and linearly defined by $\bs{\xi}_0$) as control points. Define the function $\BEZC:[0,T]\to\R^n$:
\begin{equation}
\label{eqn:bezpolyvec}
    \BEZC(\tau) = \begin{bmatrix} \BEZc(\tau) & \BEZc^{(1)}(\tau) & \cdots & \BEZc^{(n-1)}(\tau) \end{bmatrix}^\top.
\end{equation}
There exists a matrix\textsuperscript{\ref{HandDfootnote}}
$\mb D\in\R^{2n\times 2n}$ such that for any two vectors $\x_0,\x_1\in\R^n$, the unique B\'ezier curve $\BEZc$ of order $2n-1$ satisfying $\BEZC(0) = \x_0$ and $\BEZC(T) = \x_1$ with a vector of control points $\Bez_0\in\R^{2n}$ is given by:  
\begin{align}
\label{eqn:bezbndcond}
    \Bez^\top_0 =  \begin{bmatrix} \x_0^\top & \x_1^\top \end{bmatrix}\mb{D}^{-1}.
\end{align}

The following result shows how a sequence of points may be used to construct a set of B\'{e}zier curves that constitute a dynamically admissible trajectory for \eqref{eqn:openloop_state}: 

\begin{lemma}
\label{lem:bezdynadmistraj}
Let $N\in\mathbb{N}$, $\underline{t}\in\R_{\geq 0}$, and define $\overline{t}=\underline{t}+NT$ . For $k = 0,\ldots,N$, consider a collection of points $\{\mb{x}_k\}$
with $\mb{x}_k\in\R^n$ and define $t_k\in\R_{\geq 0}$ as $t_k = \underline t + kT$. For $k=0,\ldots,N-1$, let $\BEZc_k:[0,T]\to\R$ be a B\'{e}zier curve of order $2n-1$ with control points $(\Bez_k)_0 = \begin{bmatrix}(\bez_k)_{0,0} & \ldots & (\bez_k)_{0,2n-1}\end{bmatrix}^\top\in\R^{2n}$ given by:
\begin{equation}
\label{eqn:bezsetctrlpts}
    (\bs{\xi}_k)_0^\top = \begin{bmatrix} \mb{x}_{k}^\top & \mb{x}_{k+1}^\top \end{bmatrix}\mb{D}^{-1}.
\end{equation}
Defining the functions $\BEZC_k:[0,T]\to\R^n$ as in \eqref{eqn:bezpolyvec}, we have that the function $\mb{x}_d:\tintc\to\R^n$ defined as:
\begin{align}
\label{eqn:dyn_admissible_traj}
\mb{x}_d(t) &= \BEZC_k\left(t-t_k\right),\, t\in\left[t_k,t_{k+1} \right), \nonumber \\
\mb{x}_d(\overline{t}) &= \mb{x}_N,
\end{align}
is a dynamically admissible trajectory for the system \eqref{eqn:openloop_state}.
\end{lemma}

We note that the preceding result reduces planning of an (infinite dimensional) continuous time trajectory to planning a finite sequence of points. This aligns with planning dynamically admissible trajectories online in a multi-rate approach. While other classes of functions (such as general polynomials) may similarly be used to construct dynamically admissible trajectories for \eqref{eqn:openloop_state}, the motivation for using B\'{e}zier curves lies in the convex hull relationship between the curve $\BEZC_k$ and the control points $(\Bez_k)_0,\ldots,(\Bez_k)_{n-1}$. More precisely, for $i=0,\ldots,2n-1$ denote:
\begin{equation}
    (\bs{\zeta}_k)_i \triangleq \begin{bmatrix}(\bez_k)_{0,i}&\ldots& (\bez_k)_{n-1,i}\end{bmatrix}^\top\in\R^n.
\end{equation}
The points $(\Bez_k)_j$ can be viewed as the control points in time for the curve $\BEZc_k^{(j)}$, while $(\bs{\zeta}_k)_i$ reflects the control points for the curve $\BEZC_k$ realized in state space. This enables the following:

\begin{fact}[\cite{kamermans2020primer} \S4]
\label{fct:cvxhull}
We have that $\BEZC_k(\tau)\in\textrm{conv}(\{(\bs{\zeta}_k)_i\})$ for all $\tau\in[0,T]$. 
\end{fact}

\begin{figure}[t]
    \centering
    \includegraphics[width=\linewidth]{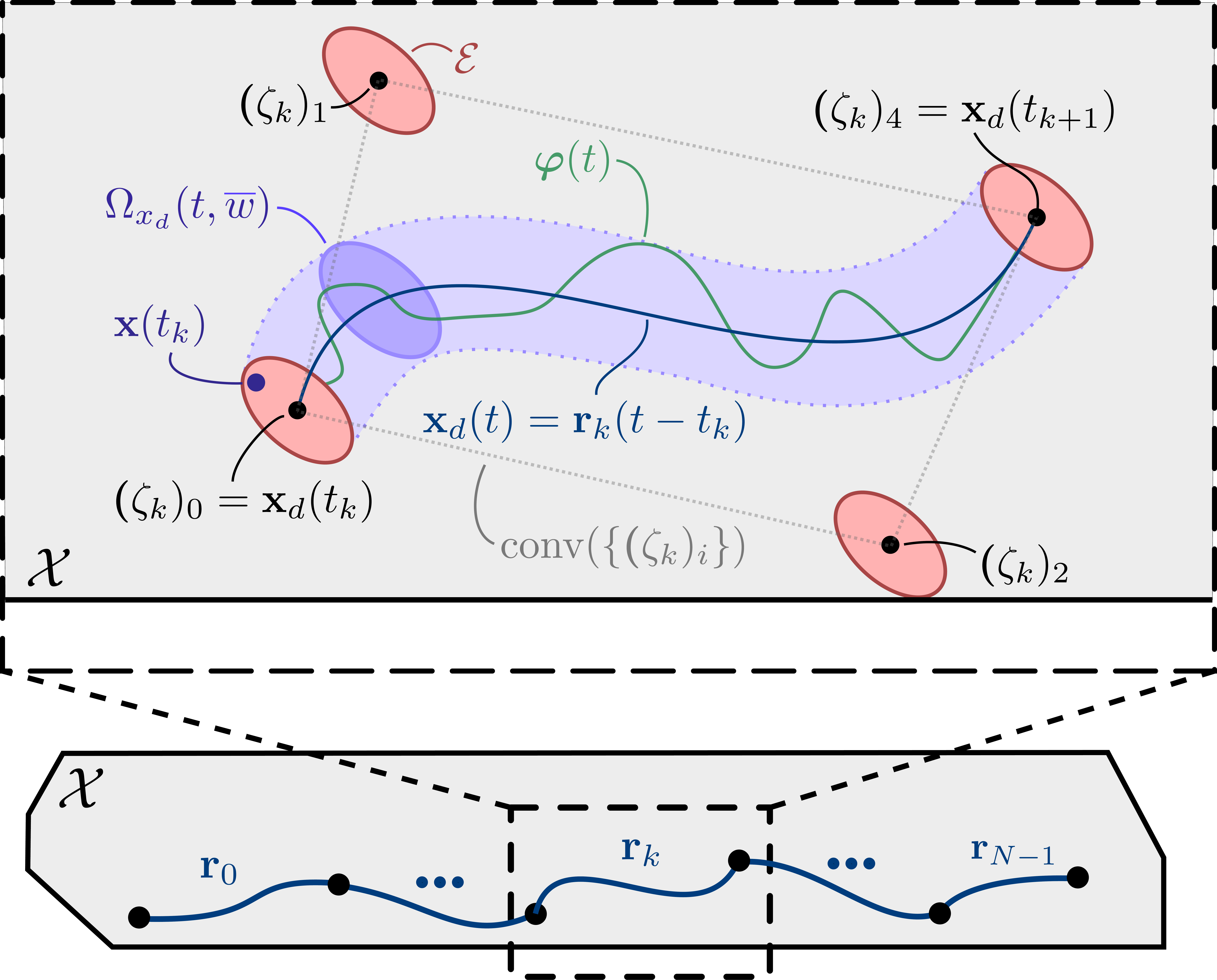}
    \caption{A depiction of the proposed method, where the control points of the B\'ezier curve are constraint tightened by the size of the robust invariant tube coming from the low-level controller.}
    \label{fig:BoundDepiction}
    \vspace{-5mm}
\end{figure}

We may immediately use this property to establish the following result regarding state constraints:

\begin{lemma}
\label{lem:robust_hull}
Define the convex, compact set $\ErrorSet \subseteq \R^n$ as:
\begin{equation}
    \ErrorSet = \{ \mb{v}\in\R^n ~|~ \mb{v}^T \mb P \mb{v} \leq \gamma \overline w^2\}.
\end{equation}
If $(\BEZ_k)_i \in \mathcal{X} \ominus \ErrorSet$ for $i =0,\ldots,2n-1$ and $k=0,\ldots,N-1$, then we have that $\Omega_{\mb{x}_d}(t,\overline{w})\subseteq \mathcal{X}$ for all $t\in\tintc$.
\end{lemma}

This result states that by constraining the B\'{e}zier curve control points, we can ensure the evolution of the system under the low-level controller satisfies state constraints. The requirement that  $(\BEZ_k)_i \in \mathcal{X} \ominus \ErrorSet$ can be expressed as an affine inequality constraint as in the following result:

\begin{lemma}
\label{lem:state_bound} 
We have that for $j=1,\ldots,q$:
\begin{align}
    (\BEZ_k)_i\in\mathcal{X} \ominus \E \Leftrightarrow \mb L_j^\top(\BEZ_k)_i \le \ell_j-\sqrt{\gamma\overline{w}^2\mb L^\top_j \mb P^{-1}\mb L_j}.
\end{align}
\end{lemma}

\section{Input Constraints}
\label{sec:input}

In this section, we present the second main contribution of this work. We show how the structure of a low-level tracking controller can be used to place requirements on a dynamically admissible trajectory $\mb{x}_d$ to ensure input constraint satisfaction.

We will make the following assumption regarding input constraints for the system:

\begin{assumption}
\label{ass:inputpolytope}
The input constraint set $\mathcal{U}\subset\R$ is given by $\mathcal{U}=[-u_{\rm max}, u_{\rm max}]$ for some $u_{\rm max}\in\mathbb{R}_{>0}$.
\end{assumption}

Neither of the controllers $k_{\mb{x}_d}^{\rm fbl}$ or $k_{\mb{x}_d}^{\rm clf}$ are necessarily required to take values in the set $\mathcal{U}$. Thus, satisfying input constraints may require violating the inequality constraint in \eqref{eqn:cntrllyapreq}, potentially invalidating the claim that $\bs{\varphi}(t)\in\RobustInvariant_{\mb{x}_d}(t,\overline{w})$ for all $t\in\tintc$. To address this limitation, knowledge of how much control action is required by the controller to track the reference trajectory under disturbances should be incorporated when synthesizing $\mb{x}_d$.

To this end, we state the following definitions. For $\alpha,\beta\in\R_{\geq 0}$, define the matrix $\mb{M}_{\alpha,\beta}\in\mathbb{S}_{\succeq 0}^2$ and the functions $\mb{N}_{\alpha,\beta}:\mathcal{X}\to\R^2_{\geq 0}$ and $\Gamma_{\alpha,\beta}:\mathcal{X}\to\R_{\geq 0}$ as:
\begin{align}
    \mb{M}_{\alpha,\beta} &= \bs{\pi}_{\rm PSD}\left(\begin{bmatrix}2\alpha\beta & \beta \\ \beta & 0 \end{bmatrix} \right), \\
    \mb{N}_{\alpha,\beta}(\overline{\mb{x}}) &= \begin{bmatrix} 2\alpha\beta\overline{e}+\alpha\vert g(\overline{\mb{x}})^{-1}\vert+\beta\Vert\mb{K}\Vert_2\overline{e} \\
    \vert g(\overline{\mb{x}})^{-1}\vert + \beta\overline{e} \end{bmatrix}, \\
    \Gamma_{\alpha,\beta}(\overline{\mb{x}}) &=\overline{e}(\beta\overline{e}+\vert g(\overline{\mb{x}})^{-1} \vert)(\alpha+\Vert\mb{K}\Vert_2),
\end{align}
where $\bs{\pi}_{\rm PSD}:\mathbb{S}^2\to\mathbb{S}^2_{\succeq 0}$ denotes the projection from symmetric matrices to symmetric positive semidefinite matrices, $\overline{e} \triangleq \sqrt{\gamma\overline{w}^2 / \lambda_{\rm min}(\mb P)}$, and $\mb{K}$ is defined in \eqref{eqn:FBL_CL_gain}. Given these definitions, we state one of our main results:


\begin{theorem}
\label{thm:fbl_bound}
There exists constants $\underline{\alpha},\underline{\smash{\beta}}\in\R_{\geq 0}$ such that if $\alpha\geq\underline{\alpha}$, $\beta\geq\underline{\smash{\beta}}$, and $\Omega_{\mb{x}_d}(t,\overline{w})\subseteq\mathcal{X}$ for all $t\in[\underline{t},\overline{t}]$, then for any collection of points $\{\overline{\mb{x}}_k\}$ with $\overline{\mb{x}}_k\in\mathcal{X}$ for $k = 0,\ldots,N-1$, we have that for all $t\in[t_k,t_{k+1})$:
\begin{align}
    \label{eqn:M_eq}
    \| k_{\mb{x}_d}^{\textnormal{fbl}}(\state,t)\|_2 \le   
    \frac{1}{2}\bs \sigma_{\mb{x}_d}(t)^\top &{\mb{M}}_{\alpha,\beta}\bs \sigma_{\mb{x}_d}(t)  \\+& {\mb{N}}_{\alpha,\beta}(\overline{\mb{x}}_k)^\top \bs \sigma_{\mb{x}_d}(t) + {\Gamma}_{\alpha,\beta}(\overline{\mb{x}}_k), \nonumber
\end{align}
for all $\mb{x}\in \Omega_{\mb{x}_d}(t,\overline{w})$, where $\bs \sigma_{\mb{x}_d}:\tintc\to\R_{\geq 0}^2$ is defined as:
\begin{align}
   \bs \sigma_{\mb{x}_d}(t) &= \begin{bmatrix}
    \|\mb{x}_d(t)-\overline{\state}_k\|_2 \\
    \|\dot{x}_d^n(t)- f(\overline{\state}_k)\|_2
    \end{bmatrix}, \quad t\in[t_k,t_{k+1}).  \label{eqn:s_def}
\end{align}
\end{theorem}

This result is motivated by the key observation that the upper bound achieved in \eqref{eqn:M_eq} is convex in the quantity $\bs \sigma_{\mb{x}_d}(t)$ for each $t\in\tintc$, such that the constraint:
\begingroup\makeatletter\def\f@size{9}\check@mathfonts
\begin{align}
    \frac{1}{2}\bs \sigma_{\mb{x}_d}(t)^\top {\mb{M}}_{\alpha,\beta}\bs \sigma_{\mb{x}_d}(t) + {\mb{N}}_{\alpha,\beta}(\overline\state_k)^\top \bs \sigma_{\mb{x}_d}(t) + {\Gamma}_{\alpha,\beta}(\overline\state_k) &\le u_{\rm max}. \label{eqn:M_ineq_desired}
\end{align}
\endgroup
is a convex quadratic inequality constraint in the quantity $\bs \sigma_{\mb{x}_d}(t)$. Given that the function $\bs \sigma_{\mb{x}_d}$ is defined by B\'{e}zier control points, we seek to translate this constraint into one on control points.

\begin{remark}
We note that the proof of Theorem \ref{thm:fbl_bound} establishes the existence of values of $\underline{\alpha}$ and $\underline{\smash{\beta}}$ through Lipschitz properties of the dynamics. In practice, it may be difficult to compute these values, and they may not necessarily be the minimum values for which this result holds. Moreover, choosing very large values of $\alpha$ and $\beta$ may lead to conservative behavior, as the constraint in \eqref{eqn:M_ineq_desired} will constrain the dynamically admissible trajectory $\mb{x}_d$ to a small neighborhood of $\overline{\mb{x}}_k$. These issues are not unexpected, as the challenge of input constraint satisfaction for general nonlinear systems is known to be quite difficult. Instead, with this result we seek to highlight an important monotonic structural property of the system that permits a well-posed and practical approach for achieving input constraint satisfaction. In particular, one may begin with small values of $\alpha$ and $\beta$ and increase them until the closed-loop nonlinear system meets input constraints. We will demonstrate this type of procedure in Section \ref{sec:sim}. 
\end{remark}

Before relating Theorem \ref{thm:fbl_bound} to the B\'{e}zier control points defining $\mb{x}_d$, we state the following lemma:
\begin{lemma}
\label{lem:eq_bound}
For any $\x\in\mathbb{R}^n$, we have that:
\begin{align}
\|\BEZC_k(\tau) - \x\|_2 &\le {\sup}_{i}\|(\BEZ_k)_i-\x\|_2, \\
\|\BEZc_k^{(n)}(\tau) - f(\x)\|_2 &\le {\sup}_{i}\|(\bez_k)_{n,i}-f(\x)\|_2,
\end{align}
for all $\tau\in[0,T]$. 
\end{lemma}

With this result, we now state one of our main results for tractably enforcing input bounds:

\begin{lemma}
\label{lem:input_bounds}
If given a collection of points $\{\overline{\mb{x}}_k\}$ with $\overline{\mb{x}}_k\in\mathcal{X}$ for $k = 0,\ldots,N-1$, there exists $\mb{s}_k\in\R^2_{\geq 0}$ such that:
\begin{align}
    \label{eqn:s_ineq}
  \begin{bmatrix}\|(\BEZ_k)_i - \overline{\mb{x}}_k\|_2 \\ \|(\bez_k)_{n,i}- f(\overline{\mb{x}}_k)\|_2\end{bmatrix} &\le {\mb s}_k,  \\ \label{eqn:t_ineq}
    \frac{1}{2}\mb s_k^\top {\mb{M}}_{\alpha,\beta}\mb s_k + {\mb{N}}_{\alpha,\beta}(\overline{\state}_k)^\top \mb s_k + {\Gamma}_{\alpha,\beta}(\overline{\state}_k) &\le u_{\rm max},
\end{align}
for $i=0,\ldots,2n-1$ and $k=0,\ldots,N-1$, then we have that the inequality \eqref{eqn:M_ineq_desired} is satisfied with $\bs \sigma_{\mb{x}_d}(t)$ defined as in \eqref{eqn:s_def} for all $t\in\tintc$. 
\end{lemma}
A consequence of this result is that for sufficiently high values of $\alpha$ and $\beta$, meeting the conditions of Lemma \ref{lem:input_bounds} implies $\| k_{\mb{x}_d}^{\rm fbl}(\mb{x},t)\|_2 \leq u_{\rm max}$ for all $t\in\tintc$ and $\mb{x}\in\Omega_{\mb{x}_d}(t,\overline{w})$. Moreover, the constraint \eqref{eqn:s_ineq} is a second-order cone constraint and the constraint \eqref{eqn:t_ineq} is a convex quadratic constraint (which may be reformulated as a second-order cone constraint, see the appendix), and thus they may be incorporated into a convex program for determining B\'{e}zier control points. Lastly, we state the following corollary relating bounds on $k_{\mb{x}_d}^{\rm fbl}$ and $k_{\mb{x}_d}^{\rm clf}$:

\begin{corollary}
\label{cor:kclfbd}
If the function $k^{\rm fbl}_{\mb{x}_d}$ is bounded as in \eqref{eqn:M_eq} for all $t\in[t_k,t_{k+1})$ and $\mb{x}\in\Omega_{\mb{x}_d}(t,\overline{w})$, then we have that:
\begin{align}
    \label{eqn:M_eqclf}
    \| k_{\mb{x}_d}^{\textnormal{clf}}(\state,t)\|_2 \le   
    \frac{1}{2}\bs \sigma_{\mb{x}_d}(t)^\top &{\mb{M}}_{\alpha,\beta}\bs \sigma_{\mb{x}_d}(t)  \\+& {\mb{N}}_{\alpha,\beta}(\overline{\mb{x}}_k)^\top \bs \sigma_{\mb{x}_d}(t) + {\Gamma}_{\alpha,\beta}(\overline{\mb{x}}_k), \nonumber
\end{align}
for all $t\in[t_k,t_{k+1})$ and all $\mb{x}\in\Omega_{\mb{x}_d}(t,\overline{w})$.
\end{corollary}

\section{Multi-Rate Control Architecture}
\label{sec:multirate}

Utilizing the developments presented in the previous sections, we now construct a multi-rate control architecture which iteratively produces dynamically admissible trajectories for the system \eqref{eqn:openloop_state} and tracks them with the low-level controller designed in Section \ref{sec:lowlevel}. Importantly, by achieving robustness to disturbances with the low-level controller, the trajectory planning algorithm can reason about a disturbance-free system.


\subsection{Model Predictive Control}
In this section we establish how to compute the collection of points $\{\mb{x}_k\}$ used to define $\mb{x}_d$ in Lemma \ref{lem:bezdynadmistraj} while meeting the desired constraints on the B\'{e}zier control points. Consider a collection of points $\{\overline\state_{k}\}$ with $\overline\state_{k}\in\mathcal{X}$ and $\{\overline u_{k}\}$ with $\overline{u}_{k}\in\R$ for $k=0,\ldots N-1$. To incorporate information about the system dynamics when synthesizing $\mb{x}_d$ as in Lemma~\ref{lem:bezdynadmistraj}, we will use linearizations of the system dynamics~\eqref{eqn:openloop_state} around these collections of points. This approximation of the dynamics will provide constraints on sequential state points (and the corresponding B\'{e}zier control points as defined by \eqref{eqn:bezsetctrlpts}) defining $\mb{x}_d$. 
We neglect the disturbances $\mb{w}$ in this approximation as the low-level controller rejects these disturbances and provides a robust invariant set around $\mb{x}_d$. More precisely, consider a linear, temporal discretization of \eqref{eqn:openloop_state}:
\begin{align}
    \label{eqn:linearizations}
    \state_{k+1} &= \A(\overline\state_{k}, \overline u_{k}) \state_{k} + \B(\overline\state_{k}) u_{k} + \C(\overline\state_{k}, \overline u_{k}),
\end{align}
where $\A:\mathcal{X}\times\R\to\R^{n\times n}, \B:\mathcal{X}\to\R^n$, and $\C:\mathcal{X}\times\R\to\R^n$ come from linearizing and taking the exact temporal discretization\footnote{See the appendix for a formula for these linearizations and discretizations.} (with sample period $T$) of the dynamics in \eqref{eqn:openloop_state}. For notational simplicity let us define:
\begin{align}
    \mb A_k \triangleq \mb A(\overline\state_k, \overline u_k),~~\mb B_k \triangleq \mb B(\overline\state_k),~~\mb C_k \triangleq \mb C(\overline\state_k, \overline u_k).
\end{align}
Given these, let us denote the state at a time $t\in\R_{\geq 0}$ by $\mb{x}(t)$. Building upon the previous two sections, we propose a Finite Time Optimal Control Problem (FTOCP):
\begin{subequations}
\begin{alignat}{2}
    \hspace{-1mm}\min_{\substack{u_{k}, \state_{k}\\ \mb s_{k}, \Bez_{k}}} \quad & \sum_{k=0}^{N-1} h(\state_{k},u_{k})  + J(\state_{N}) \label{eqn:FTOCP} \tag{FTOCP}\\
\textrm{s.t.}~ \quad & \state_{k+1} = \A_{k}\state_{k} + \B_{k}u_{k} + \C_{k}, \label{eqn:mpclindyn} \\
  & \state_{0} \in \state(t) \oplus \mathcal{E}, \label{eqn:mpcic}\\
  & \state_{N} = \mb{0}, \label{eqn:mpcterminal}\\ 
  & (\Bez_k) = \begin{bmatrix} \x_{k}^\top &\x_{k+1}^\top \end{bmatrix}\mb{D}^{-1},\label{eqn:mpcbezdef} \\
  & (\BEZ_k)_i \in \mathcal{X} \ominus \E, \hspace{23.7mm} \forall i \in \mathcal{I} \label{eqn:mpccnvxhull}\\
  & \begin{bmatrix}\|(\BEZ_k)_i - \overline{\state}_{k}\|_2 \\ \|(\bez_k)_{n,i}- f(\overline{\state}_{k})\|_2\end{bmatrix} \le \mb s_{k}, \hspace{5mm} \forall i\in\mathcal{I} \label{eqn:mpcconic}\\
  & \frac{1}{2}\mb s_{k}^\top \mb{M}_{\alpha,\beta} \mb s_{k} + {\mb N}_{\alpha,\beta}(\overline \state_k)^\top \mb s_{k} \nonumber \\ & \qquad\qquad\qquad\qquad\qquad + {\Gamma}_{\alpha,\beta}(\overline \state_k) \le u_{\rm max}, \label{eqn:mpcquadratic}
  \end{alignat}
\end{subequations}
where $h:\mathcal{X}\times\R\to\R_{\geq 0}$ is a convex stage cost, $J:\mathcal{X}\to\R_{\geq 0}$ is a convex terminal cost, and $\mathcal{I} = \{0,\ldots,2n-1\}$.
The constraint in \eqref{eqn:mpclindyn} requires that the sequence of discrete points defining $\mb{x}_d$ satisfy a linear, discrete time approximation of the system dynamics. The constraint in \eqref{eqn:mpcic} requires that the beginning of $\mb{x}_d$ is close to the current state $\mb{x}(t)$, such that $\mb{x}(t)\in\Omega(t,\overline{w})$ as required by Lemma \ref{lem:lowleveliss}. The constraint in \eqref{eqn:mpcterminal} requires the end of $\mb{x}_d$ to be placed at origin. The constraints in \eqref{eqn:mpcbezdef}-\eqref{eqn:mpcquadratic} relate the discrete points $\mb{x}_k$ to B\'{e}zier control points, and consequently the continuous trajectory $\mb{x}_d$ tracked by the low-level controller. Note that as in Fact \ref{fct:cvxhull}, the coefficients $(\bs{\xi})_k$ and $(\bs{\zeta}_k)_i$ are linearly related for $i=0,\ldots,2n-1$, a constraint implicitly assumed in \eqref{eqn:FTOCP}. If $h$ and $J$ are positive definite quadratic functions, \eqref{eqn:FTOCP} is a second-order cone program (SOCP), which can be efficiently solved via standard solvers \cite{mosek}. 

\begin{remark}
Note that we do not explicitly enforce input constraints on the decision variables $u_k$. Instead, constraints are induced on these decision variables through the linear dynamics constraint \eqref{eqn:mpclindyn} and the constraints on the B\'{e}zier coefficients in \eqref{eqn:mpcbezdef} and \eqref{eqn:mpcconic}-\eqref{eqn:mpcquadratic}. Moreover, these constraints ensure that the low-level controller will satisfy input constraints as desired. 
\end{remark}

\subsection{The Multi-Rate Architecture}

\begin{algorithm}[b!]
  \caption{$u$ = C-MPC($\state, t$)\label{alg:mpc-fl}}
  \begin{algorithmic}[1]
    \If {$t\in\mathcal{T}=\cup_{\mpcindex=0}^\infty\{\mpcindex T\}$}
        \State Compute $\{\textbf{Lin}_{k|\mpcindex}\}$ in \eqref{eqn:linearizations} about $\{\overline{\x}_{k|\mpcindex}\}$ and $\{\overline u_{k|\mpcindex}\}$;
        \State Solve \eqref{eqn:FTOCP} with $\{\textbf{Lin}_{k|\mpcindex}\}$;
        \If {\eqref{eqn:FTOCP} is infeasible}
        \State $\{\textbf{Lin}_{k|\mpcindex}\} \leftarrow \{\textbf{Lin}_{1|\mpcindex-1},\ldots,\textbf{Lin}_{N-1|\mpcindex-1},\textbf{Lin}_{O}\};$
        \State Solve \eqref{eqn:FTOCP}  with $\{\textbf{Lin}_{k|\mpcindex}\}$;
        \EndIf
        \State $\{\overline\x_{k|\mpcindex+1}\} \leftarrow \{\x_{1|\mpcindex}^*,\ldots,\x^*_{N-1|\mpcindex},\x^*_{N|\mpcindex}\}$;
        \State $\{\overline u_{k|\mpcindex+1}\} \leftarrow \{ u_{1|\mpcindex}^*,\ldots, u_{N-1|\mpcindex}^*,0\}$;
    \EndIf
    \State Calculate $\x_d | \mpcindex$ from $\{\mb x_{k|\mpcindex}^*\}$, as in \eqref{eqn:bezsetctrlpts}--\eqref{eqn:dyn_admissible_traj};
    \State \Return $u = k^{\textrm{clf}}_{\state_{d}|\mpcindex}(\x, t)$;
  \end{algorithmic}
\end{algorithm}

We now present the multi-rate architecture that integrates the low-level controller design posed in Section \ref{sec:lowlevel} with the preceding trajectory planner encoded in \eqref{eqn:FTOCP}.

We first recall the role $T$ plays in dynamically admissible trajectories synthesized through B\'{e}zier curves as in Lemma \ref{lem:bezdynadmistraj}, as well as its role as a sampling period for the temporal discretization established in \eqref{eqn:linearizations}. Let us denote $\mathcal{T}=\cup_{\mpcindex=0}^\infty\{\mpcindex T\}$. This set serves to index the discrete points in time (separated by $T$) at which a dynamically admissible trajectory for the system will be replanned by solving the \eqref{eqn:FTOCP}. The multi-rate architecture is initialized at time $t=0$ with collections of points $\{\overline{\mb{x}}_{k|0}\}$ and $\{\overline{u}_{k|0}\}$ with $\overline{\mb{x}}_{k|0}\in\mathcal{X}$ and $\overline{u}_{k|0}\in\R$ for $k=0,\ldots,N-1$. Let us denote the linearized and discretized dynamics computed around these collections by $\{\textbf{Lin}_{k|0}\} = \{(\mb{A}_{k|0},\mb{B}_{k|0},\mb{C}_{k|0})\}$.
\begin{assumption}
Given an initial condition $\mb{x}(0)\in\mathcal{X}$, \eqref{eqn:FTOCP} is feasible using $\{\textbf{Lin}_{k|0}\}$. 
\end{assumption}

We now describe our multi-rate framework as summarized in Algorithm \ref{alg:mpc-fl}. As in Line 1, let $t\in\mathcal{T}$ such that $t=iT$ for some $i\in\mathbb{Z}$. In Line 2, the linearized and discretized dynamics are computed around the collections $\{\overline{\mb{x}}_{k|i}\}$ and $\{\overline{u}_{k|i}\}$, and are denoted by $\{\textbf{Lin}_{k|i}\} = \{(\mb{A}_{k|i},\mb{B}_{k|i},\mb{C}_{k|i})\}$. In Line 3 these dynamics are used to solve the \eqref{eqn:FTOCP} using the state at the current time, $\mb{x}(t)$, in \eqref{eqn:mpcic}. If the \eqref{eqn:FTOCP} is feasible, it returns collections of points $\{\mb{x}^*_{k|i}\}$ with $\mb{x}^*_{k|i}\in\mathcal{X}$ for $k=0,\ldots, N$ and $\{u^*_{k|i}\}$ with $u^*_{k|i}\in\R$ for $k=0,\ldots, N-1$. If \eqref{eqn:FTOCP} is infeasible, in Line 5 we set the linearized and discretized dynamics $\{\textbf{Lin}_{k|i}\}$ to the previous linearization shifted by one and appending the linearization and discretization around the origin, denoted $\textbf{Lin}_O = (\mb{A}(\mb{0},0),\mb{B}(\mb{0}),\mb{C}(\mb{0},0))$. In Line 6 we solve \eqref{eqn:FTOCP} and similarly return collections of points $\{\mb{x}^*_{k|i}\}$ and $\{u^*_{k|i}\}$. As we will show in Theorem \ref{thm:TheBigCahoona}, our assumption about feasibility at time $t=0$ will ensure that switching to this set of linearizations will always ensure \eqref{eqn:FTOCP} is feasible. In Line 8--9 the collection $\{\mb{x}^*_{k|i}\}$ is shifted and the collection $\{u^*_{k|i}\}$ is shifted and appended with $0$ to define collections $\{\overline{\mb{x}}_{k|i+1}\}$ and $\{\overline{u}_{k|i+1}\}$ used for linearization and discretization in the next iteration. In Line 11 the collection $\{\mb{x}^*_{k|i}\}$ is then used to define a dynamically admissible trajectory $\mb{x}_d|i$ as in Lemma \ref{lem:bezdynadmistraj}, which yields a corresponding low-level controller  $k^{\textrm{clf}}_{\state_{d}|\mpcindex}$ that defines the output of our algorithm. We may view our algorithm as a time-varying controller that yields a closed-loop system \eqref{eqn:closedloop_state}. Importantly, our algorithm ensures state and input constraints are satisfied as the continuous time system evolves under this controller, as stated in the following theorem: 

\begin{figure}[t!]
    \centering
    \includegraphics[width=\linewidth]{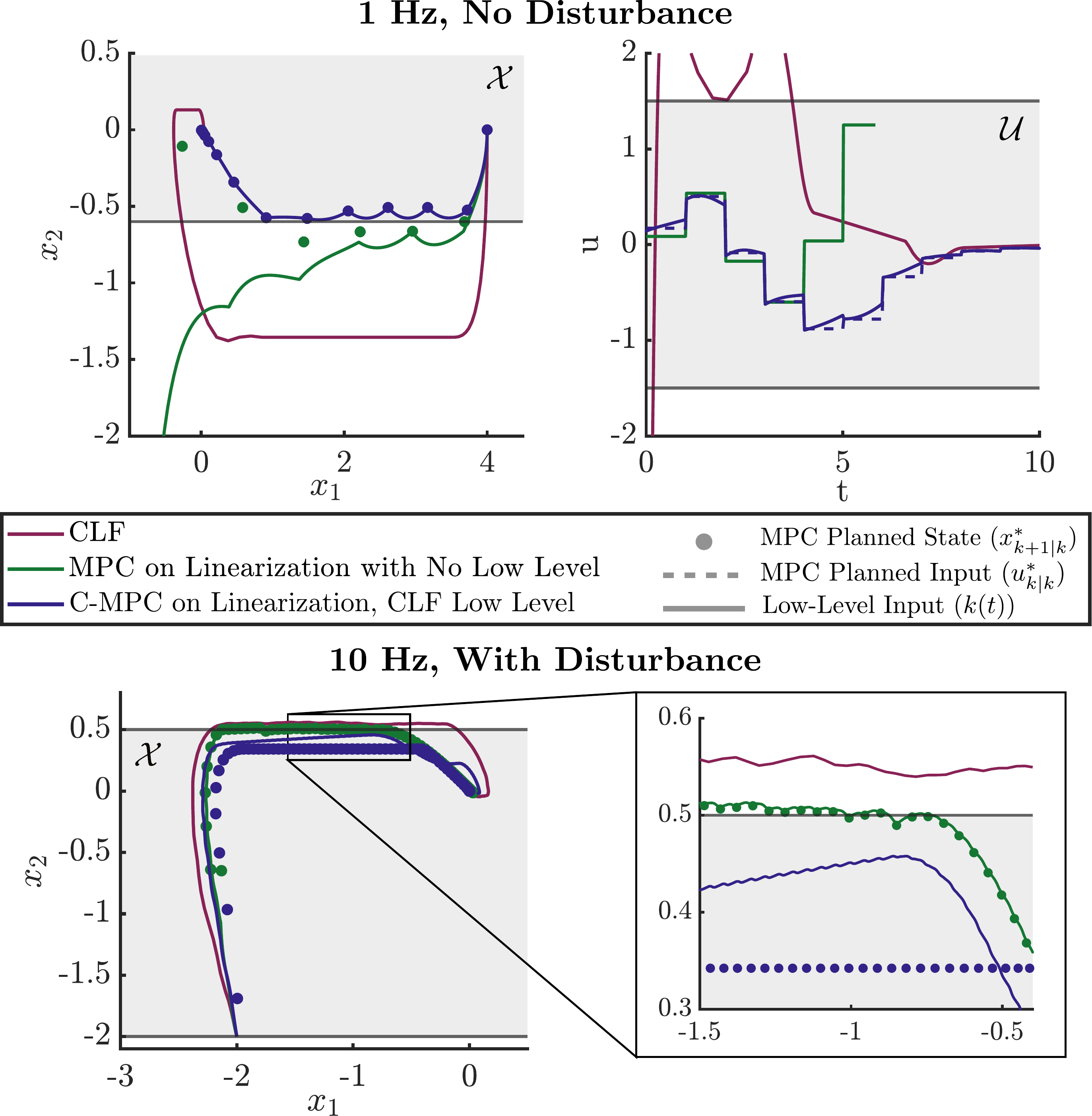}
    \caption{Comparison of three control methods: only using a low level controller (CLF), applying MPC with no low-level controller, and applying the proposed C-MPC with a CLF at the low-level. In both scenarios, just using the low-level or mid-level controller separately yields both state and input violation.}
    \label{fig:comparison}
\end{figure}

\begin{theorem}
\label{thm:TheBigCahoona}
Suppose that $\alpha\geq\underline{\alpha}$ and $\beta\geq\underline{\smash\beta}$ are such that $\Gamma_{\alpha,\beta}(\mb{0})\leq u_{\rm max}$. Let \eqref{eqn:FTOCP} be defined with $\alpha$ and $\beta$, and consider the closed-loop system \eqref{eqn:closedloop_state} with a feedback controller given by C-MPC in Algorithm \ref{alg:mpc-fl} and a disturbance signal satisfying $\Vert\mb{w}\Vert_{\infty}\leq\overline{w}$. If $\mb{0}\in\mathcal{X}\ominus\ErrorSet$ and \eqref{eqn:FTOCP} is feasible at $t_0=0$ with initial condition $\mb{x}(0)\in\mathcal{X}$, then C-MPC is well-defined for all time, and the closed-loop system \eqref{eqn:closedloop_state} satisfies state and input constraints.
\end{theorem}

\section{Simulation}
\label{sec:sim}



\begin{figure}[t!]
    \centering
    \includegraphics[width=\linewidth]{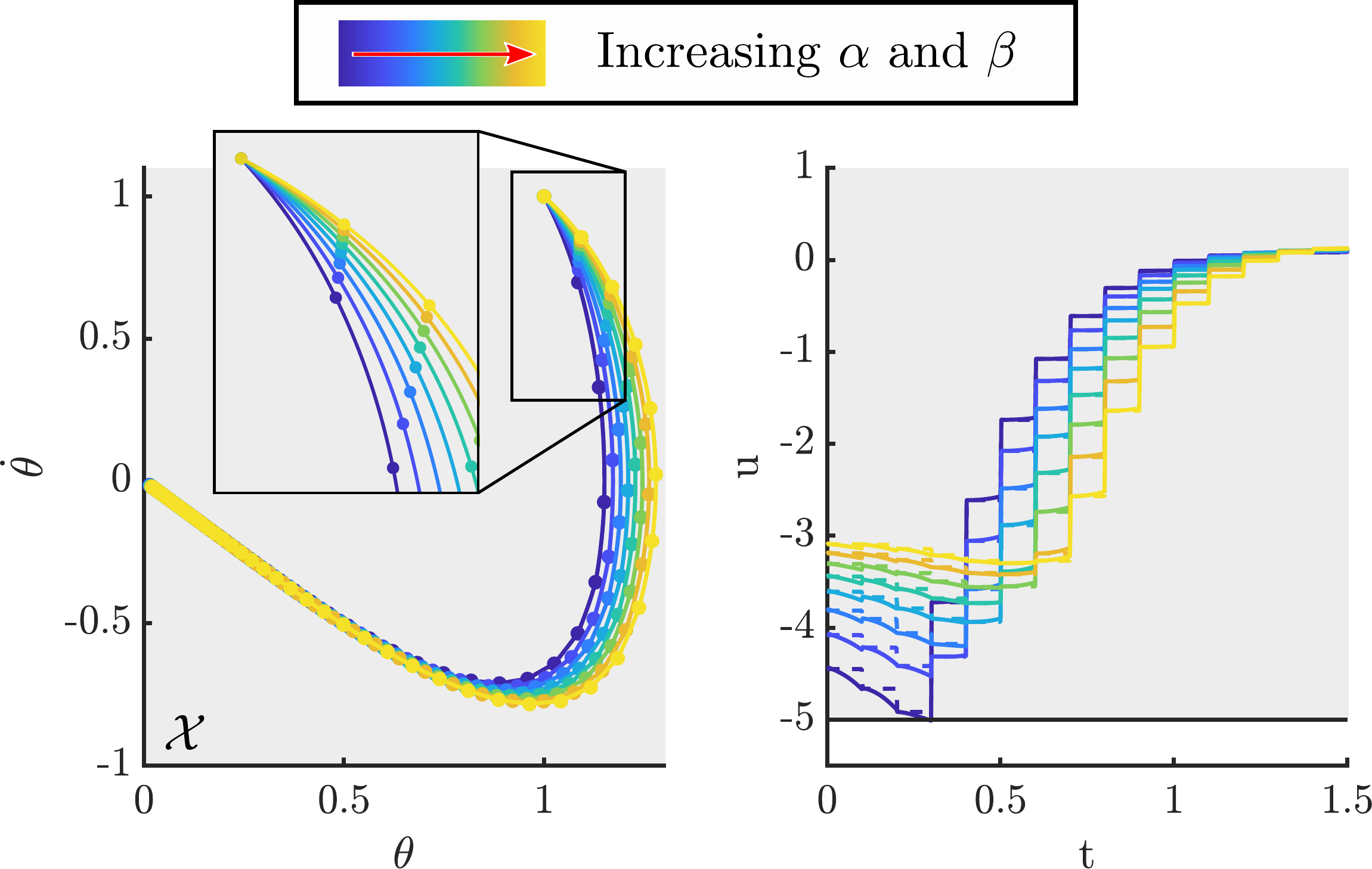}
    \caption{The proposed C-MPC for increasing user parameter values $\alpha$ and $\beta$. Notice that as the parameters increase, the planned MPC points become spatially closer so as to reduce the linearization error, and in doing so the deviation of the low-level controller from the planned control input decreases.}
    \label{fig:increasing_gain}
\end{figure}

We consider the following nonlinear system in simulation:
\begin{align*}
    \begin{bmatrix}\dot x_1 \\ \dot x_2\end{bmatrix} = \begin{bmatrix}0&1\\0&0\end{bmatrix}\begin{bmatrix}x_1\\x_2\end{bmatrix} + \begin{bmatrix}0\\\sin(x_1)+x_2^3\end{bmatrix} + \begin{bmatrix}0\\1\end{bmatrix}u + \begin{bmatrix}w_1(t)\\ w_2(t)\end{bmatrix}.
\end{align*}
The goal is to drive the system to the origin while satisfying state and input constraints for all time. Fig.~\ref{fig:comparison} demonstrates that at different time scales, both with and without added disturbances, using only either a low-level or mid-level controller results in state and/or input violation, whereas the proposed combined approach is able to satisfy both for all time. Fig.~\ref{fig:increasing_gain} shows the behavior of the system for increasing values of $\alpha$ and $\beta$. As the parameter values increase, the planned MPC points become closer to reduce deviation from the linearization points, and in doing so the deviation of the low-level controller from the planned input $u_k$ decreases as the system evolves from $\mb{x}^*_k$ to $\mb{x}^*_{k+1}$. Simulation code is provided at \cite{csomay_shanklin_cmpc_2022}.


\section{Conclusion and Future Work}
In conclusion, we have presented a multi-rate control architecture for nonlinear systems that utilizes MPC in conjunction with B\'{e}zier curves to iteratively plan continuous time trajectories that are tracked using Control Lyapunov Function based controllers. Our approach allows us to ensure that the low-level controller satisfies state and input constraints as it tracks the desired trajectory. We believe there are a number of meaningful directions for future work. First, in the pursuit of a truly multi-\textit{rate} scheme, the low-level CLF control design could be adapted to the sampled-data setting \cite{taylor2021sampled}. Next, our work uses the origin as the terminal set, but developing constructive approaches to synthesizing terminal sets using the ideas in \cite{marcucci2021shortest} could greatly improve the feasible domain of our method. Lastly, we believe that the challenge of underactuation and unstable \textit{zero-dynamics} may be best approached through a joint planning and low-level control mindset, and believe our work serves as a first step in this direction \cite{kohler2021constrained}.

\bibliographystyle{bibtex/myIEEEtran} 
\bibliography{main}

\clearpage \section{Appendix}
\subsection{Construction of $\mb{H}$:}
Let $r$ be a B\'ezier curve of order $p$ defined as in \eqref{eqn:bez}:
\begin{align*}
    \BEZc(\tau) = \sum_{i=0}^{p}\bez_{i} z_i(\tau),
\end{align*}
with control points given by the elements of $\bs{\xi} = \begin{bmatrix}\xi_0 & \cdots & \xi_p\end{bmatrix}^\top$.
The derivative of this curve is given by:
\begin{equation*}
    \dot{r}(\tau) = \sum_{i=0}^{p}\bez_{i} \dot{z}_i(\tau),
\end{equation*}
which we may equivalently express as \cite[\S 13]{kamermans2020primer}:
\begin{align*}
    \dot\BEZc(\tau) = \frac{1}{T} \sum_{i=0}^{p-1}p(\bez_{i+1}-\bez_i) z_i(\tau) \triangleq \frac{1}{T} \sum_{i=0}^{p-1}\vartheta_i z_i(\tau),
\end{align*}
with $\vartheta_i \triangleq p(\xi_{i+1}-\xi_i)$ for $i=0,\ldots,p-1$. We observe that $\dot{r}$ is a B\'{e}zier curve of order $p-1$ with control points $\vartheta_i/T$. We may increase the order of $\dot{r}$ by one (making it a B\'{e}zier curve of order $p$) by the following transformation \cite[\S 12]{kamermans2020primer}:
\begin{align*}
    \dot\BEZc(\tau) = \frac{1}{T} \sum_{i=0}^{p}\left(\frac{(p-i)\vartheta_i+i\vartheta_{i-1}}{p}\right) z_i(\tau) \triangleq \frac{1}{T}\sum_{i=0}^{p} \omega_i z_i(\tau),
\end{align*}
where $\vartheta_{-1}=\vartheta_p\triangleq0$. Thus $\dot{r}$ is a B\'{e}zier curve of order $p$ with control points $\omega_i/T$. Noting that $\omega_i$ is a linear function of $\xi_{i-1},\xi_i$, and $\xi_{i+1}$, we may rewrite $\dot{r}$ as:
\begin{align*}
    \dot{\BEZc}(\tau) = \frac{1}{T}\Bez^\top {\underbrace{(\mb S^\top \mb R^\top)}_{\mb H}}  \z (\tau),
\end{align*}
where $\mb S \in \mathbb{R}^{p\times p+1}$ and $\mb R \in \mathbb{R}^{p+1\times p}$ are defined by:
\begin{align*}
    \mb S_{ii} &= -p, ~~\mb S_{i,i+1} = p, \\
    \mb R_{ii} &= \frac{p+1-i}{p}, ~~ \mb R_{i+1,i} = \frac{i}{p},
\end{align*}
for $i=1,\ldots,p$ with all other entries zero \cite{kamermans2020primer}. The matrix $\mb S$ corresponds to the differentiation of $r$, and the matrix $\mb R$ corresponds to increasing the order of the curve by one. Furthermore, we may reapply this transformation an arbitrary number of times to produce higher-order derivatives:
\begin{equation*}
    {\BEZc}^{(j)}(\tau) = \frac{1}{T^j} \Bez^\top \mb H^j  \z (\tau).
\end{equation*}

\subsection{Construction of $\mb D$}
Let us denote:
\begin{align*}
    \mb{x}_0 =&~ \begin{bmatrix} x_{0,0} & \cdots & x_{0,n-1}\end{bmatrix}^\top, \\ \mb{x}_1 =&~ \begin{bmatrix} x_{1,0} & \cdots & x_{1,n-1}\end{bmatrix}^\top.
\end{align*}
Consider the set of equality constraints on the boundary of the B\'ezier curve:
\begin{align*}
    \BEZc^{(j)}(0) &= x_{0,j},~~~j=0,\ldots,n-1\\
    \BEZc^{(j)}(T) &= x_{1,j},~~~j=0,\ldots,n-1.
\end{align*}
Substituting in the definition of a B\'ezier curve results in:
\begin{align*}
    \Bez_{j}^\top\mb z(0) &= x_{0,j},~~~j=0,\ldots,n-1\\
    \Bez_{j}^\top\mb z(T)&= x_{1,j},~~~j=0,\ldots,n-1.
\end{align*}
As all of the control points are linear in $\Bez_0,$ we can again reformulate this as:
\begin{align*}
    \Bez^\top_0\frac{1}{T^j}\mb H^j\mb z(0) &= x_{0,j},~~~j=0,\ldots,n-1\\
    \Bez^\top_0\frac{1}{T^j}\mb H^j\mb z(T) &= x_{1,j},~~~j=0,\ldots,n-1.
\end{align*}
From this, we can construct a collection of linear equality constraints:
\begin{align*}
    \Bez^\top_0\underbrace{\begin{bmatrix}\mb D_0 & \mb D_1 \end{bmatrix}}_{\mb D} = \begin{bmatrix}\state_0^\top &\mb \state_1^\top\end{bmatrix},
\end{align*}
with the matrices $\mb D_0\in\R^{2n\times n}$ and $\mb D_1\in\R^{2n\times n}$ defined as:
\begin{align*}
\mb D_0&= \begin{bmatrix}\frac{1}{T^0}\mb H^0\z(0) & \cdots & \frac{1}{T^{n-1}}\mb H^{n-1}\z(0)\end{bmatrix},\\
\mb D_1 &= \begin{bmatrix}\frac{1}{T^0}\mb H^0\z(T) & \cdots & \frac{1}{T^{n-1}}\mb H^{n-1}\z(T) \end{bmatrix}.
\end{align*}


\subsection{Proof of Lemma \ref{lem:bezdynadmistraj}:}
\begin{proof}
Let $k\in\{0,\ldots,N-1\}$. As each function $\BEZc_k^{(j)}$, $j=0,\ldots,n-1$, is a B\'ezier polynomial, the function $\BEZC_k$ is continuously differentiable on the interval $(0,T)$ and the respective one-sided limits of the derivative exist at $0$ and $T$. The definition of the B\'{e}zier control points in \eqref{eqn:bezsetctrlpts} implies that:
\begin{equation}
    \BEZC_k(T) = \begin{cases} \BEZC_{k+1}(0) \quad &\textrm{if~}k\in\{0,\ldots,N-2\}, \\ \mb{x}_N \quad &\textrm{if~}k=N-1. \end{cases} \nonumber
\end{equation}
Thus the function $\mb{x}_d$ is continuous on $\tintc$, which with the previous differentiability properties, implies it is piecewise continuously differentiable on $\tintc$.
Next, observe that:
\begin{equation}
\dot{\mb{x}}_d(t) = \dot{\BEZC}_k\left(t-t_k\right),\, t\in\left(t_k, t_{k+1}\right), \notag
\end{equation}
where $\dot{\BEZC}_k:(0,T)\to\R^n$ is given by:
\begin{align*}
    \dot{\BEZC}_k(\tau) = \begin{bmatrix} \BEZc_k^{(1)}(\tau) & \cdots & \BEZc_k^{(n)}(\tau) \end{bmatrix}^\top,
\end{align*}
with $r_k^{(j)}$ defined as in \eqref{eqn:bezjderiv}. This may be rewritten as:
\begin{align*}
    \dot{\BEZC}_k(\tau) &= \begin{bmatrix} \mb 0&\mb I\\0&\mb{0}^\top\end{bmatrix}\BEZC_k(\tau)+\begin{bmatrix}\mb 0\\ \BEZc_k^{(n)}(\tau) \end{bmatrix}.
\end{align*}
Thus we have that:
\begin{align*}
    \dot{\mb{x}}_d(t) &= \begin{bmatrix} \mb 0&\mb I\\0&\mb{0}^\top\end{bmatrix}\mb{x}_d(t)+\begin{bmatrix}\mb 0\\ 1 \end{bmatrix}r_k^{(n)}(t-t_k),
\end{align*}
for $t\in\left(t_k,t_{k+1}\right)$. Defining the function $u_d:\tintc\to\R$ as:
\begin{equation}
    u_d(t) = g(\mb{x}_d(t))^{-1}(-f(\mb{x}_d(t))+r^{(n)}_k(t-t_k)), \nonumber
\end{equation}
for $t\in[t_k, t_{k+1})$, the continuity of $f,g$ on $\R^n$, the continuity of $r_k^{(n)}$ on $[0,T]$, and the fact $g(\mb{x})\neq 0$ for all $\mb{x}\in\R^n$, implies $u_d$ is piecewise continuous. Moreover, we have that:
\begin{equation}
    \dot{\x}_d(t) = \f(\x_d(t))+\g(\x_d(t))u_d(t), \nonumber
\end{equation}
for almost all $t\in\tintc$. Thus $\mb{x}_d$ is a dynamically admissible trajectory for the system \eqref{eqn:openloop_state}.
\end{proof}

\subsection{Proof of Lemma \ref{lem:robust_hull}:}
\begin{proof}
Because $\mathcal{X}$ and $\ErrorSet$ are convex, their Minkowski difference, $\mathcal{X}\ominus\ErrorSet$, is also convex. As such, if $(\BEZ_k)_i\in\mathcal{X}\ominus\ErrorSet$ for $i=0,\ldots,2n-1$, then $\text{conv}(\{(\BEZ_k)_i\})\subseteq\mathcal{X}\ominus\ErrorSet$. As in Fact \ref{fct:cvxhull}, the convex hull property of B\`ezier curves implies that $\BEZC_k(\tau)\in\textrm{conv}(\{(\bs{\zeta}_k)_i\})$ for all $\tau\in[0,T]$. Thus we have that $\BEZC_k(\tau)\in\mathcal{X}\ominus\ErrorSet$ for all $\tau\in[0,T]$, implying that for any time $t\in \tintc$ we have that $\mb \state_d(t)$ as defined in \eqref{eqn:dyn_admissible_traj} is contained in $\mathcal{X}\ominus\mathcal{E}$. Therefore, $\mb \state_d(t)\oplus \mathcal{E}=\Omega_{\mb{x}_d}(t,\overline w)\subseteq \mathcal{X}$, as desired.
\end{proof}

\subsection{Proof of Lemma \ref{lem:state_bound}:}
\begin{proof}
First, suppose that $(\bs{\zeta}_k)_i\in\mathcal{X}\ominus \E$ and let $j\in\{1,\ldots,q\}$. We then have that:
\begin{align}
    \mb{L}_j^\top((\bs{\zeta}_k)_i+\mb{v}) \leq {\ell}_j,~~\forall \mb{v} \in \ErrorSet. \nonumber
\end{align}
Equivalently, we have that:
\begin{equation}
 \mb{L}_j^\top(\bs{\zeta}_k)_i 
     \leq {\ell}_j - \sup_{\mb{v}\in\ErrorSet} \mb{L}_j^\top\mb{v}. \nonumber
\end{equation}
Noting that $\E=\{\mb v~|~\mb v^\top\mb P\mb v \le \gamma\overline{w}^2\}$ is a convex set as $\mb{P}$ is positive definite, taking the Lagrangian yields:
\begin{align*}
  \sup_{\mb{v}\in\ErrorSet} \mb{L}_j^\top\mb{v} &= \adjustlimits \inf_{\lambda\in\R_+} \sup_{\mb{v}\in\R^n} \mb L_j^\top \mb v - \lambda (\mb v^\top \mb P\mb v-\gamma\overline{w}^2), \\ &\triangleq \adjustlimits \inf_{\lambda\in\R_+}   \sup_{\mb{v}\in\R^n} \mathcal{L}(\mb{v},\lambda).
\end{align*}
The stationarity conditions implies that: 
\begin{align}
\label{eqn:stationarity}
    \nabla_{\mb v} \mathcal{L}(\mb v^*,\lambda^*) = \mb L_j - 2\lambda^* \mb P \mb v^* = 0     \implies \mb v^* = \frac{1}{2\lambda^*}\mb P^{-1}\mb L_j. \nonumber
\end{align}
Substituting this expression for $\mb{v}^*$ into the Lagrangian yields the dual problem:
\begin{equation}
   \sup_{\mb{v}\in\ErrorSet} \mb{L}_j^\top\mb{v} =  \inf_{\lambda\in\R_{+}}  \frac{1}{4\lambda}\mb{L}_j^\top\mb{P}^{-1}\mb{L}_j+\lambda\gamma\overline{w}^2. \nonumber
\end{equation}
The stationarity condition yields:
\begin{equation}
    \lambda^* = \frac{1}{2}\sqrt{\frac{\mb{L}_j^\top\mb{P}^{-1}\mb{L}_j}{\gamma\overline{w}^2}}, \nonumber
\end{equation}
From this we arrive at:
\begin{equation}
    \sup_{\mb{v}\in\ErrorSet}\mb{L}_j^\top\mb{v} = \sqrt{\gamma\overline{w}^2\mb{L}_j^\top\mb{P}^{-1}\mb{L}_j}, \nonumber
\end{equation}
whereby we then know that:
\begin{equation}
    \mb{L}_j^\top(\bs{\zeta}_k)_i \leq \ell_j - \sqrt{\gamma\overline{w}^2\mb{L}_j^\top\mb{P}^{-1}\mb{L}_j},  \nonumber
\end{equation}
as desired. 

Second, let $j\in\{1,\ldots,N\}$, and suppose that:
\begin{equation}
    \mb{L}_j^\top(\bs{\zeta}_k)_i \leq \ell_j - \sqrt{\gamma\overline{w}^2\mb{L}_j^\top\mb{P}^{-1}\mb{L}_j},  \nonumber
\end{equation}
Then for $\mb{w}\in\ErrorSet$, we have that:
\begin{equation}
    \mb{L}_j^\top((\bs{\zeta}_k)_i+\mb{w}) \leq \mb{L}_j^\top(\bs{\zeta}_k)_i+\sup_{\mb{v}\in\ErrorSet}\mb{L}_j^\top\mb{v} \leq \ell_j, \nonumber
\end{equation}
following from our previous evaluation of the supremum. Thus we have $(\bs{\zeta}_k)_i+\mb{w}\in\mathcal{X}$, and since $\mb{w}$ and $j$ were arbitrary, we have $(\bs{\zeta}_k)_i\in\mathcal{X}\ominus\ErrorSet$.
%
%
\end{proof}
\subsection{Proof of Theorem \ref{thm:fbl_bound}}
\begin{proof} 
\phantom{\qedhere}
Let $k\in\{0,\ldots,N-1\}$, let $t\in[t_k,t_{k+1})$ and let $\mb{x}\in\Omega_{\mb{x}_d}(t,\overline{w})$. For notational simplicity, let $g^\dagger:\R^n\to\R$ be defined as $g^\dagger(\mb{x}) = g(\mb{x})^{-1}$. We first bound the feed-forward input defined in \eqref{eqn:feedforward} as follows:
\begin{align}
    \|k_{\mb{x}_d}^{\textrm{ff}}(\mb{x},t)\|_2 &= \left\|g^{\dagger}(\mb{x})\mathcal{F}_{\mb{x}_d}(\mb{x},t)\right\|_2, \nonumber \\
    & \leq  \left\|g^{\dagger}(\mb{x})\right\|_2\left\|f(\mb{x})-\dot{x}_d^n(t)\right\|_2. \nonumber 
\end{align}
with $\mathcal{F}_{\mb{x}_d}$ defined in \eqref{eqn:calFxd}. Given this, we have that:
\begin{align}
   \hspace{-0.2 cm} \|k_{\mb{x}_d}^{\textrm{fbl}}(\mb{x},t)\|_2 &\leq \|k_{\mb{x}_d}^{\textrm{fbl}}(\mb{x},t) - k_{\mb{x}_d}^{\textrm{ff}}(\mb{x},t)\|_2 + \|k_{\mb{x}_d}^{\textrm{ff}}(\mb{x},t)\|_2, \label{eqn:fbl_bound} \\
    &\leq \|g^{\dagger}(\mb{x})\mb{K}^\top\mb{e}_{\mb{x}_d}(\mb{x},t)\|_2 + \|k_{\mb{x}_d}^{\textrm{ff}}(\mb{x},t)\|_2, \notag \\
    &\leq \left\|g^{\dagger}(\mb{x})\right\|_2\left(\|\mb{K}\|_2 \overline{e}+\left\|f(\mb{x})-\dot{x}_d^n(t)\right\|_2\right). \notag
\end{align}
As the set $\mathcal{X}$ is compact, we have that $f$ and $g^\dagger$ are Lipschitz continuous on $\mathcal{X}$ (as $g$ is Lipschitz continuous and non-zero) with Lipschitz constants $L_f,L_{g^\dagger}\in\R_{\geq 0}$, respectively. We continue by observing that:
\begin{align}
    \|g^{\dagger}(\mb{x})\|_2 &\leq
    \|g^{\dagger}(\mb{x})-g^{\dagger}(\mb{x}_d(t))\|_2+\|g^{\dagger}(\mb{x}_d(t))-g^{\dagger}(\overline{\mb{x}}_k)\|_2 \notag \\
    &\hspace{4mm}+\|g^{\dagger}(\overline{\mb{x}}_k)\|_2, \notag \\ 
    &\le L_{g^{\dagger}}(\|\mb e_{\mb{x}_d}(\state,t)\|_2 + \|\state_d(t) - \overline{\state}_k\|_2)+\|g^{\dagger}(\overline{\mb{x}}_k)\|_2, \notag \\ 
    &\le L_{g^{\dagger}}(\overline{e} + \|\state_d(t) - \overline{\state}_k\|_2)+\|g^{\dagger}(\overline{\mb{x}}_k)\|_2. \nonumber
\end{align}
Similarly, we have that:
\begin{align}
    \|f(\state)-\dot{x}^n_d(t))\|_2 \le L_{f}(\overline{e} &+ \|\state_d(t)  -\overline{\state}_k\|_2) \nonumber \\
    &+ \|f(\overline\state_k)-\dot{x}^n_d(t))\|_2. \notag
\end{align}
The previous bounds allow us to construct a matrix $\mb{M}\in \mathbb{S}^{2}$ and functions $\mb N:\R^n\to\mathbb{R}_{\geq 0}^{2}$, and $\Gamma:\R^n\to\mathbb{R}_{\geq 0}$ defined as:
\begin{align*}
    \mb{M} &= \begin{bmatrix} 2L_{g^{\dagger}}L_{f} & L_{g^{\dagger}}\\ L_{g^{\dagger}}&0\end{bmatrix},\\
    \mb N(\overline\state_k) &= \begin{bmatrix} 2L_{g^{\dagger}}L_{f}\overline{e} + L_{f}\|g^{\dagger}(\overline{\state}_k)\|_2+L_{g^{\dagger}}\|\mb{K}\|_2\overline{e} \\ \|{g^{\dagger}(\overline{\state}_k)}\|_2 +L_{g^{\dagger}}\overline{e} \end{bmatrix}, \\
    \Gamma(\overline\state_k) &= \overline{e}(L_{g^{\dagger}}\overline{e}+\|g^{\dagger}(\overline{\state}_k)\|_2)(L_{f}+\|\mb{K}\|_2),
\end{align*}
such that:
\begin{align*}
    \|k_{\mb{x}_d}^{\textrm{fbl}}(\mb{x},t)\|_2 &\leq \frac{1}{2} \bs{\sigma}_{\mb{x}_d}(t)^\top \mb{M}\bs{\sigma}_{\mb{x}_d}(t) + \mb N(\overline\state_k)^\top \bs{\sigma}_{\mb{x}_d}(t) + \Gamma(\overline\state_k).
\end{align*}
where:
\begin{align*}
   \bs \sigma_{\mb{x}_d}(t) &= \begin{bmatrix}
    \|\mb{x}_d(t)-\overline{\state}_k\|_2 \\
    \|\dot{x}_d^n(t)- f(\overline{\state}_k)\|_2
    \end{bmatrix}, \quad t\in[t_k,t_{k+1}).  \label{eqn:s_def}
\end{align*}
Let $\underline{\alpha} = L_f$ and $\underline{\smash\beta} = L_{g^\dagger}$. We can then see that if both $\alpha\geq\underline{\alpha}$ and $\beta\geq\underline{\smash\beta}$, then:
\begin{equation*}
    \mb{N}_{\alpha,\beta}(\overline{\mb{x}}_k) \geq   \mb{N}(\overline{\mb{x}}_k),
\end{equation*}
where the inequality is element-wise, and:
\begin{equation*}
    \Gamma_{\alpha,\beta}(\overline{\mb{x}}_k) \geq   \Gamma(\overline{\mb{x}}_k).
\end{equation*}
As the elements of $\mb{N}$ and $\mb{N}_{\alpha,\beta}$ are non-negative, we have:
\begin{equation*}
    \mb{N}_{\alpha,\beta}(\overline{\mb{x}}_k)^\top\mb{v} \geq   \mb{N}(\overline{\mb{x}}_k)^\top\mb{v},
\end{equation*}
for any $\mb{v}\in\R^{2}_{\geq 0}$. Given the definition of $\bs{\sigma}_{\mb{x}_d}$ (with non-negative elements by definition of a norm), we thus have:
\begin{align*}
    \|k_{\mb{x}_d}^{\textrm{fbl}}(\mb{x},t)\|_2 \leq \frac{1}{2} \bs{\sigma}_{\mb{x}_d}(t)^\top & \mb{M}\bs{\sigma}_{\mb{x}_d}(t)  \notag \\+& \mb N_{\alpha,\beta}(\overline\state_k)^\top \bs{\sigma}_{\mb{x}_d}(t) + \Gamma_{\alpha,\beta}(\overline\state_k).
\end{align*}
We next observe that if both $\alpha\geq\underline{\alpha}$ and $\beta\geq\underline{\smash\beta}$, then:
\begin{align*}
    \frac{1}{2}\mb{v}^\top\mb{M}\mb{v} & = L_{g^\dagger}L_fv_1^2 +L_{g^\dagger}v_1v_2 \\ & \leq \alpha\beta v_1^2 + \beta v_1v_2 \triangleq \frac{1}{2}\mb{v}^\top\widetilde{\mb{M}}_{\alpha,\beta}\mb{v},
\end{align*}
for all $\mb{v} = \begin{bmatrix} v_1 & v_2 \end{bmatrix}^\top \in \R^2_{\geq 0}$, where:
\begin{equation*}
    \widetilde{\mb{M}}_{\alpha,\beta} = \begin{bmatrix} 2\alpha\beta & \beta \\ \beta & 0 \end{bmatrix}.
\end{equation*}
It can be seen that the matrix $\widetilde{\mb{M}}_{\alpha,\beta}$ will have both a positive and a negative eigenvalue for any $\alpha,\beta\in\R_{\geq 0}$, and thus using $\widetilde{\mb{M}}_{\alpha,\beta}$ directly in \eqref{eqn:M_ineq_desired} will yield a non-convex constraint for an optimization program. To resolve this, we will project $\widetilde{\mb{M}}_{\alpha,\beta}$ onto the positive semidefinite cone to get $\mb{M}_{\alpha,\beta}$, such that:
\begin{align*}
    \mb M_{\alpha,\beta}  \triangleq \pi_{\textrm{PSD}}(\widetilde{\mb{M}}_{\alpha,\beta}),  =  \lambda_1(\widetilde{\mb{M}}_{\alpha,\beta}) \mb v_1(\widetilde{\mb{M}}_{\alpha,\beta})\mb v_1(\widetilde{\mb{M}}_{\alpha,\beta})^\top,
\end{align*}
where $\lambda_1(\widetilde{\mb{M}}_{\alpha,\beta})$ is the positive eigenvalue of $\widetilde{\mb{M}}_{\alpha,\beta}$, and:
\begin{equation}
\label{eqn:Mabeigvec}
    \mb v_1(\widetilde{\mb{M}}_{\alpha,\beta}) = \frac{1}{\sqrt{1+\lambda_1(\widetilde{\mb{M}}_{\alpha,\beta})^2}}\begin{bmatrix} \lambda_1(\widetilde{\mb{M}}_{\alpha,\beta}) \\ 1 \end{bmatrix},
\end{equation}
is the corresponding unit eigenvector. By construction, we have that $\mb M_{\alpha,\beta} \succeq \widetilde{\mb{M}}_{\alpha,\beta}$, and thus we may conclude that:
\begin{equation*}
     \frac{1}{2}\mb{v}^\top\mb{M}\mb{v} \leq \frac{1}{2}\mb{v}^\top\widetilde{\mb{M}}_{\alpha,\beta}\mb{v} \leq \frac{1}{2}\mb{v}^\top\mb{M}_{\alpha,\beta}\mb{v},
\end{equation*}
for all $\mb{v}\in\R^2_{\geq 0}$. Thus we can conclude that:
\begin{align*}
    \|k_{\mb{x}_d}^{\textrm{fbl}}(\mb{x},t)\|_2 \leq \frac{1}{2} \bs{\sigma}_{\mb{x}_d}(t)^\top & \mb{M}_{\alpha,\beta}\bs{\sigma}_{\mb{x}_d}(t)  \notag \\+& \mb N_{\alpha,\beta}(\overline\state_k)^\top \bs{\sigma}_{\mb{x}_d}(t) + \Gamma_{\alpha,\beta}(\overline\state_k). \qed
\end{align*}
\end{proof}

\begin{remark}
The projection of the matrix $\widetilde{\mb M}_{\alpha,\beta}$ onto the positive semidefinite cone is a relaxation of the problem in that it will shrink the set of feasible dynamically admissible trajectories. In doing so, it provides a tractable way for guaranteeing that input bounds are met. Importantly, this is a type of ``minimal" relaxation as the projection onto the positive semidefinite cone is the closest matrix that yields a convex inequality constraint.
\end{remark}


\subsection{Proof of Lemma \ref{lem:eq_bound}:}

\begin{proof}
Let $\mb{x}\in\R^n$. The convex hull property of B\'{e}zier curves implies that for any $\tau\in[0,T]$, we may write:
\begin{equation*}
    \BEZC_k(\tau) = \sum_{i=0}^{2n-1}\lambda_i(\tau)(\bs{\zeta}_k)_i,
\end{equation*}
where $\lambda_i(\tau)\geq 0$ and $\sum_{i=0}^{2n-1} \lambda_i(\tau) = 1$. Thus we have that:
\begin{align*}
    \Vert \BEZC_k(\tau) - \mb{x} \Vert_2 
    & = \left\Vert  \sum_{i=0}^{2n-1}\lambda_i(\tau)\left((\bs{\zeta}_k)_i - \mb{x}\right)  \right\Vert_2,
\end{align*}
using the fact that $\sum_{i=0}^{2n-1} \lambda_i(\tau) = 1$. Moving the norm inside the sum, we have that:
\begin{align*}
    \Vert \BEZC_k(\tau) - \mb{x} \Vert_2 & \leq   \sum_{i=0}^{2n-1}\left\Vert\lambda_i(\tau)\left((\bs{\zeta}_k)_i - \mb{x}\right)  \right\Vert_2, \\
    & = \sum_{i=0}^{2n-1}\lambda_i(\tau)\left\Vert(\bs{\zeta}_k)_i - \mb{x}  \right\Vert_2, \\ 
\end{align*}
as $\lambda_i(\tau)\geq 0$. We may further conclude that:
\begin{align*}
    \Vert \BEZC_k(\tau) - \mb{x} \Vert_2 & \leq  \sum_{i=0}^{2n-1}\lambda_i(\tau)\sup_i\left\Vert(\bs{\zeta}_k)_i - \mb{x}  \right\Vert_2, \\ 
    & = \sup_i\left\Vert(\bs{\zeta}_k)_i - \mb{x}  \right\Vert_2, 
\end{align*}
as desired. To establish the second part, we begin by noting that $\BEZc^{(n)}$ is a B\'{e}zier curve of order $2n-1$, such that:
\begin{align*}
    \Vert \BEZc^{(n)}(\tau)-f(\mb{x}) \Vert_2 = \left\Vert \sum_{i= 0}^{2n-1}(\xi_k)_{n,i} z_i(\tau) - f(\mb{x})\right\Vert,
\end{align*}
Noting that:
\begin{equation*}
    \sum_{i= 0}^{2n-1} z_i(\tau) = 1
\end{equation*}
for all $\tau\in[0,T]$ (see \cite{farin2002curves}), we have that:
\begin{align*}
    \Vert \BEZc^{(n)}(\tau)-f(\mb{x}) \Vert_2 &= \left\Vert \sum_{i= 0}^{2n-1}((\xi_k)_{n,i}-f(\mb{x})) z_i(\tau) \right\Vert, \\
    & \leq  \sum_{i= 0}^{2n-1}\Vert((\xi_k)_{n,i}-f(\mb{x})) z_i(\tau) \Vert, \\
     & \leq  \sum_{i= 0}^{2n-1}\vert(\xi_k)_{n,i}-f(\mb{x})\vert\vert z_i(\tau) \vert,
\end{align*}
Noting that $z_i(\tau)\geq 0$ for all $\tau\in[0,T]$ and $i\in\{0,\ldots,2n-1\}$, we then have that:
\begin{align*}
    \Vert \BEZc^{(n)}(\tau)-f(\mb{x}) \Vert_2
     & \leq  \sum_{i= 0}^{2n-1}\vert(\xi_k)_{n,i}-f(\mb{x})\vert z_i(\tau), \\  & \leq  \sum_{i= 0}^{2n-1}\sup_i\vert(\xi_k)_{n,i}-f(\mb{x})\vert z_i(\tau),
     \\  & \leq  \sup_i\vert(\xi_k)_{n,i}-f(\mb{x})\vert \sum_{i= 0}^{2n-1} z_i(\tau), \\
     & = \sup_i\vert(\xi_k)_{n,i}-f(\mb{x})\vert, 
\end{align*}
yielding the desired result.
\end{proof}

\subsection{Proof of Lemma \ref{lem:input_bounds}}
\begin{proof}
Assume that the inequalities in \eqref{eqn:s_ineq} and \eqref{eqn:t_ineq} hold for some $k\in\{0,\ldots,N-1\}$. From Lemma \ref{lem:eq_bound}, we have that:
\begin{align*}
    \begin{bmatrix}
    \|\BEZC_k(\tau) - \overline \state_k\|_2 \\ 
    \|\BEZc^{(n)}_k(\tau) - f(\overline \state_k)\|_2
    \end{bmatrix} \le \mb s_k,
\end{align*}
for all $\tau\in[0,T]$, where the inequality is element-wise. Given the definition of the dynamically admissible trajectory $\mb{x}_d$ in \eqref{eqn:dyn_admissible_traj}, we then have that:
\begin{align*}
    \bs\sigma_{\mb{x}_d}(t) \le \mb s_k
\end{align*}
for all $t\in[t_k,t_{k+1})$. As the elements of $\mb M_{\alpha,\beta}$ are positive (as the elements of $\mb v_1(\widetilde{\mb{M}}_{\alpha,\beta})$ in \eqref{eqn:Mabeigvec} are both positive), and the elements of $\bs{\sigma}_{\mb{x}_d}(t)$ and $\mb{s}_k$ are non-negative, we have that:
\begin{align*}
    \bs \sigma_{\mb{x}_d}(t)^\top \mb M_{\alpha,\beta} \bs\sigma_{\mb{x}_d}(t) \le
    \mb s^\top_k \mb M_{\alpha,\beta} \mb s_k.
\end{align*}
Furthermore, as the elements of $\mb N_{\alpha,\beta}(\overline \state_k)$ are non-negative, we have that:
\begin{align*}
    \mb N_{\alpha,\beta}(\overline \state_k)^\top \bs \sigma_{\mb{x}_d}(t) \le \mb N_{\alpha,\beta}(\overline \state_k)^\top \mb s_k.
\end{align*}
Therefore, we can conclude that:
\begin{align*}
    \frac{1}{2}\bs \sigma_{\mb{x}_d}(t)&^\top {\mb{M}}_{\alpha,\beta}\bs \sigma_{\mb{x}_d}(t) + {\mb{N}}_{\alpha,\beta}(\bar\state_k)^\top \bs \sigma_{\mb{x}_d}(t) + {\Gamma}_{\alpha,\beta}(\bar\state_k) \\
    &\le \frac{1}{2}\mb s_k^\top {\mb{M}}_{\alpha,\beta}\mb s_k + {\mb{N}}_{\alpha,\beta}(\bar\state_k)^\top \mb s_k + {\Gamma}_{\alpha,\beta}(\bar\state_k)\\
    &\le u_{\rm max},
\end{align*}
as enforced via \eqref{eqn:t_ineq}.
\end{proof}

\subsection{Reformulation to a SOCP}
Consider a positive semidefinite matrix $\mb M_{\alpha,\beta}\in \mathbb{S}^2_{\succeq 0}$. We may take its Cholesky decomposition, yielding:
\begin{align*}
    \mb M_{\alpha,\beta} = \mb L_{\alpha,\beta} \mb L_{\alpha,\beta}^\top,
\end{align*}
for some $\mb L_{\alpha,\beta}\in \R^{2\times 2}$. Let $\mb{s}_k\in\R^2$. We have that:
\begin{equation}
\label{eqn:qcconst}
\frac{1}{2}\mb s_k^\top {\mb{M}}_{\alpha,\beta}\mb s_k + {\mb{N}}_{\alpha,\beta}(\overline{\state}_k) \mb s_k + {\Gamma}_{\alpha,\beta}(\overline{\state}_k) \le u_{\rm max},    
\end{equation}
if and only if there exists a $\sigma_k \in \R$ such that:
\begin{align}
\label{eqn:socpconst}
    \left\|\begin{bmatrix}\mb L_{\alpha,\beta}^\top & \mb 0\\\mb 0&1\end{bmatrix} \begin{bmatrix} \mb s_k \\ \sigma_k\end{bmatrix}\right\|_2 &\le \sigma_k + \frac{1}{2}, 
\end{align}
and:
\begin{equation}
\label{eqn:auxconst}
    \sigma_k + \frac{1}{4} \le -\mb N_{\alpha,\beta}(\overline{\mb{x}}_k)^\top\mb s_k\mb -\Gamma_{\alpha,\beta} +u_{max}.
\end{equation}
To see the if direction, assume there exists a $\sigma_k\in\R$ such that \eqref{eqn:socpconst} and \eqref{eqn:auxconst} hold. We then have that:
\begin{equation*}
     \left\|\begin{bmatrix}\mb L_{\alpha,\beta}^\top & \mb 0\\\mb 0&1\end{bmatrix} \begin{bmatrix} \mb s_k \\ \sigma_k \end{bmatrix}\right\|_2^2 \le \sigma_k^2 + \sigma_k + \frac{1}{4},
\end{equation*}
which may be rewritten as:
\begin{equation*}
    \mb s_k^\top \mb L_{\alpha,\beta} \mb L_{\alpha,\beta}^\top \mb s_k + \sigma_k^2\le \sigma_k^2 + \sigma_k + \frac{1}{4}.
\end{equation*}
Using the definition of $\mb{L}_{\alpha,\beta}$, we arrive at:
\begin{equation*}
    \mb s_k^\top \mb M_{\alpha,\beta} \mb s_k \le -\mb N_{\alpha,\beta}(\overline{\state}_k)^\top s_k -\Gamma_{\alpha,\beta}(\overline{\state}_k) + u_{max}, 
\end{equation*}
and thus have:
\begin{equation*}
    \mb s_k^\top \mb M_{\alpha,\beta} \mb s_k + \mb N_{\alpha,\beta}(\overline{\state}_k)^\top\mb s_k + \Gamma_{\alpha,\beta}(\overline{\state}_k) \le u_{max}
\end{equation*}
For the only if direction, suppose that \eqref{eqn:qcconst} is satisfied, and let:
\begin{equation*}
    \sigma_k = -\mb N_{\alpha,\beta}(\overline{\mb{x}}_k)^\top\mb s_k\mb -\Gamma_{\alpha,\beta} +u_{max} - \frac{1}{4},
\end{equation*}
such that \eqref{eqn:auxconst} is satisfied. Substituting this into \eqref{eqn:qcconst} yields:
\begin{equation*}
     \mb s_k^\top \mb M_{\alpha,\beta} \mb s_k \leq \sigma_k + \frac{1}{4}.
\end{equation*}
Adding $\sigma_k^2$ to each side and using the definition of $\mb{L}_{\alpha,\beta}$ yields:
\begin{equation*}
     \mb s_k^\top \mb L_{\alpha,\beta} \mb{L}_{\alpha,\beta}^\top \mb s_k + \sigma_k^2 \leq \sigma_k^2 + \sigma_k + \frac{1}{4}.
\end{equation*}
This may be rewritten as:
\begin{equation*}
     \left\|\begin{bmatrix}\mb L_{\alpha,\beta}^\top & \mb 0\\\mb 0&1\end{bmatrix} \begin{bmatrix} \mb s_k \\ \sigma_k \end{bmatrix}\right\|_2^2 \le \left(\sigma_k+\frac{1}{2}\right)^{2}.
\end{equation*}
Taking the square root of each side yields \eqref{eqn:socpconst} as desired.

\subsection{Proof of Corollary \ref{cor:kclfbd}}
\begin{proof} \phantom{\qedhere}
Let $k\in\{0,\ldots,N-1\}$, let $t\in[t_k,t_{k+1})$, and let $\mb{x}\in\Omega_{\mb{x}_d}(t,\overline{w})$. From \eqref{eqn:lyapdecay} we know that $k_{\mb{x}_d}^{\rm fbl}(\mb{x})$ is a feasible solution to the optimization problem defining $k_{\mb{x}_d}^{\rm clf}$, and thus we may conclude:
\begin{align*}
    \frac{1}{2}\|k_{\mb{x}_d}^{\textrm{clf}}(\bs \state,t) - k_{\mb{x}_d}^{\textrm{ff}}(\bs \state,t)\|_2^2 \le  \frac{1}{2}\|k_{\mb{x}_d}^{\textrm{fbl}}(\bs \state,t) - k_{\mb{x}_d}^{\textrm{ff}}(\bs \state,t)\|_2^2.
\end{align*}
From this we have that:
\begin{align*}
    & \|k_{\mb{x}_d}^{\textrm{clf}}(\bs \state,t) - k_{\mb{x}_d}^{\textrm{ff}}(\bs \state,t)\|_2+ \|k_{\mb{x}_d}^{\textrm{ff}}(\mb{x},t)\|_2  \\ & \qquad\qquad\qquad \le  \|k_{\mb{x}_d}^{\textrm{fbl}}(\bs \state,t) - k_{\mb{x}_d}^{\textrm{ff}}(\bs \state,t)\|_2+ \|k_{\mb{x}_d}^{\textrm{ff}}(\mb{x},t)\|_2.
\end{align*}
From the triangle inequality, we have that:
\begin{align*}
    \|k_{\mb{x}_d}^{\textrm{clf}}(\bs \state,t)\|_2 \le \|k_{\mb{x}_d}^{\textrm{clf}}(\bs \state,t) - k_{\mb{x}_d}^{\textrm{ff}}(\bs \state,t)\|_2 + \|k_{\mb{x}_d}^{\textrm{ff}}(\bs \state,t)\|_2.
\end{align*}
Using this to replace the left-hand side of the inequality in \eqref{eqn:fbl_bound}, we may proceed as in the proof of Theorem \ref{thm:fbl_bound} to arrive at:
\begin{align*}
    \|k_{\mb{x}_d}^{\textrm{clf}}(\mb{x},t)\|_2 \leq \frac{1}{2} \bs{\sigma}_{\mb{x}_d}(t)^\top & \mb{M}_{\alpha,\beta}\bs{\sigma}_{\mb{x}_d}(t)  \notag \\+& \mb N_{\alpha,\beta}(\overline\state_k)^\top \bs{\sigma}_{\mb{x}_d}(t) + \Gamma_{\alpha,\beta}(\overline\state_k).\qed
\end{align*}
\end{proof}

\subsection{Linearization and Discretization:}
We can linearize the dynamics of \eqref{eqn:openloop_state} to generate a linear, continuous time representation about the point $(\overline \state_k,\overline u_k)$ as:
\begin{align*}
    \A_c(\overline \state_k, \overline u_k) &= \frac{\partial \mb f}{\partial \state}({\overline \state_k})+\frac{\partial \mb g}{\partial \state}(\overline \state_k)\overline u_k,\\ 
    \B_c(\overline \state_k) &= \mb g(\overline \state_k),\\
    \C_c(\overline \state_k, \overline u_k) &= \mb f(\overline \state_k) + \mb g(\overline \state_k)\overline u_k - \A_c(\overline \state_k, \overline u_k) \overline \state_k - \B_c(\overline \state_k) \overline u_k.
\end{align*}
We can then employ exact temporal discretization over a time interval $T$ to obtain:
\begin{align*}
    \A(\overline \state_k, \overline u_k) &= e^{\A_c(\overline \state_k, \overline u_k)T}, \\
    \B(\overline \state_k) &= \int_0^{T}e^{\mb{A}_c(\overline \state_k, \overline u_k)(T-\tau)}\B_c(\overline \state_k) d\tau,\\
    \C(\overline \state_k, \overline u_k) &= \int_0^{T}e^{\mb{A}_c(\overline \state_k, \overline u_k)(T-\tau)}\C_c(\overline \state_k, \overline u_k)d\tau.
\end{align*}

\subsection{Proof of Theorem \ref{thm:TheBigCahoona}}
\begin{proof}
Let $i\in\mathbb{Z}_{\geq 0}$. Suppose that at time $t_i=iT$ and state $\mb{x}(t_i)\in\mathcal{X}$, we have that \eqref{eqn:FTOCP} is feasible using the collections of points $\{\overline{\mb{x}}_{k|i}\}$ and $\{\overline{u}_{k|i}\}$ for $k=0,\ldots,N-1$ and the corresponding linearizations $\{\textbf{Lin}_{k|i}\}.$ Let $\{\mb{x}^{*}_{k|i}\}$ for $k=0,\ldots,N$ and $\{u^*_{k|i}\}$, $\{(\bs{\xi}^*_{k|i})\}$, and $\{\mb{s}^*_{k|i}\}$ for $k=0,\ldots,N-1$ be the collection of points composing the solution to \eqref{eqn:FTOCP}, and let $\mb{x}_{d}|i:[t_i,t_{i+1}]\to\mathcal{X}$ be the continuous reference trajectory defined as in Lemma \ref{lem:bezdynadmistraj}. Given that $\Vert\mb{w}\Vert_\infty\leq\overline{w}$, Lemma \ref{lem:lowleveliss} implies that $\bs{\varphi}(t)\in\Omega(t,\overline{w})$ for all $t\in[t_i,t_{i+1}]$. We have from Lemma \ref{lem:robust_hull} that $\Omega(t,\overline{w})\subseteq\mathcal{X}$ for all $t\in[t_i,t_{i+1}]$, implying that $\bs{\varphi}(t)\in\mathcal{X}$ for all $t\in[t_i,t_{i+1}]$. Given this, we may further conclude from Theorem \ref{thm:fbl_bound}, Corollary \ref{cor:kclfbd}, and Lemma \ref{lem:input_bounds} that:
\begin{align*}
    \|k^{\textrm{clf}}_{\mb{x}_d}(\bs{\varphi}(t), t)\|_2 \le u_{max} \implies k^{\textrm{clf}}_{\mb{x}_d}(\bs{\varphi}(t), t) \in\mathcal{U}.
\end{align*}
for all $t\in[t_i,t_{i+1}]$.

To see that our algorithm is recursively feasible (i.e, feasible at time $t_{i+1}$ given feasibility at time $t_i$), it is sufficient for us to show that \eqref{eqn:FTOCP} is feasible at the time $t_{i+1}$ with $\mb{x}(t_{i+1}) = \bs{\varphi}(t_{i+1})$ and linearizations:
\begin{equation*}
\{\textbf{Lin}_{k|i+1}\} = \{\textbf{Lin}_{1|i}, \ldots, \textbf{Lin}_{N-1|i}, \textbf{Lin}_O\},     
\end{equation*}
for $k=0,\ldots,N-1$, i.e., those calculated at time $t_i$ shifted by one index and appended with the linearization at the origin. This reflects the case in which solving \eqref{eqn:FTOCP} with the linearizations about the previous optimal solution is infeasible, so it is sufficient to check feasibility only in this case. 

To show that under such these conditions a feasible solution for \eqref{eqn:FTOCP} is given by:
\begin{align*}
    \{\widehat{\x}_{k|i+1}\} &= \left\{\mb x^*_{1|i},\ldots,\mb x^*_{N|i}, \mb 0\right\},\, &k = 0,\ldots,N,\\
    \{\widehat{u}_{k|i+1}\} &= \left\{u^*_{1|i},\ldots, u^*_{N-1|i},0\right\}, \, &k = 0,\ldots,N-1,\\
    \{\widehat{\mb s}_{k|i+1}\} &= \left\{ \mb s^*_{1|i},\ldots,\mb s^*_{N-1|i}, \mb 0\right\}, \, &k = 0,\ldots,N-1,\\
    \{(\widehat{\Bez}_{k|i+1})\} &= \left\{ (\Bez^*_{1|i}),\ldots,(\Bez^*_{N-1|i}),\mb 0\right\}, \, &k = 0,\ldots,N-1,
\end{align*}

First, we observe that:
\begin{equation}
    \mb{x}_{k+1|i}^* = \mb{A}_{k|i}\mb{x}_{k|i}^*+\mb{B}_{k|i}u_{k|i}^*+\mb{C}_{k|i}
\end{equation}
for $k=1,\ldots,N-1$ by the previous solution. Thus:
\begin{equation*}
    \widehat{\mb{x}}_{k+1|i+1} = \mb{A}_{k|i+1}\widehat{\mb{x}}_{k|i+1}+\mb{B}_{k|i+1}\widehat{u}_{k|i+1}+\mb{C}_{k|i+1},
\end{equation*}
for $k = 0,\ldots, N-2$. Noting that $\widehat{\mb{x}}_{N-1|i+1}=\mb{x}^*_{N|i}=\mb{0}$ as required by the previous solution, and the fact that $\widehat{\mb{x}}_{N|i+1}=\mb{0}$ and $\widehat{u}_{N-1|i+1}=0$, and $\mb{C}_{N-1|i}= \mb{C}(\mb{0},0)=\mb{0}$ we have that:
\begin{align*}
\widehat{\mb{x}}_{N|i+1} = \mb{A}_{N-1|i+1}\widehat{\mb{x}}_{N-1|i+1}&+\mb{B}_{N-1|i+1}\widehat{u}_{N-1|i+1}\\&+\mb{C}_{N-1|i+1}.
\end{align*}
Thus we have:
\begin{equation*}
    \widehat{\mb{x}}_{k+1|i+1} = \mb{A}_{k|i+1}\widehat{\mb{x}}_{k|i+1}+\mb{B}_{k|i+1}\widehat{u}_{k|i+1}+\mb{C}_{k|i+1},
\end{equation*}
$k = 0,\ldots, N-1$ as required in \eqref{eqn:mpclindyn}.

Next, we observe that by Lemma \ref{lem:lowleveliss}, we must have that $\bs{\varphi}(t_{i+1})\in\Omega_{\mb{x}_d|i}(t_{i+1},\overline{w})$, such that $\bs{\varphi}(t_{i+1})\in\mb{x}_d|i(t_{i+1})\oplus\ErrorSet = \mb{x}_{1|i}^*\oplus\ErrorSet$. Noting that if $\mb{v}\in\ErrorSet$, we must have that $-\mb{v}\in\ErrorSet$, and thus $\mb{x}_{1|i}^*\in\bs{\varphi}(t_{i+1})\oplus\ErrorSet$. As
$\mb{x}(t_{i+1})=\bs{\varphi}(t_{i+1})$, we have $\widehat{\mb{x}}_{k|i+1} = \mb{x}_{1|i}^*\in\mb{x}(t_{i+1})\oplus\ErrorSet$ as required by \eqref{eqn:mpcic}. As $\widehat{\mb{x}}_{N|i+1}=\mb{0}$, \eqref{eqn:mpcterminal} is satisfied.

By virtue of the previous solution, we can see that \eqref{eqn:mpcbezdef}-\eqref{eqn:mpcquadratic} are satisfied for $k=0,\ldots,N-2$ by our proposed solution. We need only show they hold for $k = N-1$. As $\widehat{\mb{x}}_{N-1|i+1} = \widehat{\mb{x}}_{N|i+1} = \mb{0}$ and $(\widehat{\bs{\xi}}_{N-1|i+1}) = \mb{0}$, we have that \eqref{eqn:mpcbezdef} is satisfied. Furthermore, we have that $(\widehat{\bs{\xi}}_{N-1|i+1}) = \mb{0}$ implies the corresponding $(\widehat{\bs{\zeta}}_{N-1|i+1})_j = \mb{0}$ for $j=0,\ldots,2n-1$. By assumption we have that $\mb{0}\in\mathcal{X}\ominus\ErrorSet$, and thus we have that \eqref{eqn:mpccnvxhull} is satisfied.

Lastly, we note that $\overline{\mb{x}}_{N-1|i+1}$ used to define the linearization $\textbf{Lin}_{N-1|i+1}$ is the origin, i.e., $\overline{\mb{x}}_{N-1|i+1}=\mb{0}$. Thus we have that the left-hand side of \eqref{eqn:mpcconic} is $\mb{0}$, and thus because $\Gamma_{\alpha,\beta}(\mb{0})\leq u_{\rm max}$, we see that $\widehat{\mb{s}}_{N-1|i+1}=\mb{0}$ satisfies \eqref{eqn:mpcconic} and \eqref{eqn:mpcquadratic}, such that our proposed solution is feasible.

\end{proof}
\begin{remark}
Note that recomputing the linearizations about the previous trajectory is not strictly necessary to ensure feasibility -- trajectories generated from \emph{any} linearization of the system will be feasible for a full-state feedback linearizeable system. However, what is payed is performance -- keeping the trajectory close to its linearization will reduce the conservativeness that the MPC program exhibits. As such, there is a conditional statement in Algorithm \ref{alg:mpc-fl}: if a feasible trajectory about the new linearizations can be found, use it, and if not, use the previous linearizations as a contingency plan to ensure feasibility. 
\end{remark}


%

\end{document}